\documentclass{aa}  

\usepackage{graphicx}
\usepackage{txfonts}
\usepackage{gensymb}
\usepackage{amsmath}
\usepackage{hyperref}
\usepackage{orcidlink}
\newcommand{\code}[1]{\texttt{#1}\xspace}
\begin{document}
   \title{Metallicities for more than 10 million stars derived from $Gaia$ BP/RP spectra \thanks{Table \ref{table:catalogue} is only available in electronic form at the CDS via anonymous ftp to cdsarc.u-strasbg.fr (130.79.128.5) or via http://cdsweb.u-strasbg.fr/cgi-bin/qcat?J/A+A/.}}

   \author{T. Xylakis-Dornbusch\, \orcidlink{0000-0002-1296-2907} \inst{1},
          N. Christlieb\, \orcidlink{0000-0002-4043-2727} \inst{1},  T.T. Hansen\, \orcidlink{0000-0001-6154-8983} \inst{2}, T.Nordlander\, \orcidlink{0000-0001-5344-8069}\inst{3,4}, K. B. Webber\, \orcidlink{0000-0002-9762-4308} \inst{5,6} \and J. Marshall\, \orcidlink{0000-0003-0710-9474} \inst{5,6}
          }

   \institute{Zentrum f\"ur Astronomie der Universit\"at Heidelberg, Landessternwarte, K\"onigstuhl 12, 69117 Heidelberg, Germany\\
              \email{txylaki@lsw.uni-heidelberg.de}
              \and
              Department of Astronomy, Stockholm University, AlbaNova University Center, SE-106 91 Stockholm, Sweden\\ 
              \and
            Research School of Astronomy and Astrophysics, Australian National University, Canberra, ACT 2611, Australia\\ 
            \and
            ARC Centre of Excellence for All Sky Astrophysics in 3 Dimensions (ASTRO 3D), Australia\\
            \and
              George P. and Cynthia Woods Mitchell Institute for Fundamental Physics and Astronomy, Texas A\&M University, College Station, TX 77843, USA \\
              \and
              Department of Physics \& Astronomy, Texas A\&M University, 4242 TAMU, College Station, TX 77843, USA \\
             }


 
  \abstract
   {The third $Gaia$ Data Release, which includes BP/RP spectra for 219 million sources, has opened a new window in the exploration of the chemical history and evolution of the Milky Way. The wealth of information encapsulated in these data is far greater than their low resolving power ($R\sim50$) at first glance would suggest, as shown in many studies. We zero in on the use of this data for the purpose of the detection of `new' metal-poor stars, which are hard to find yet essential for understanding - among other - several aspects of the origin of the Galaxy, star formation and the creation of the elements.}
   {We strive to refine a metal-poor candidate selection method which was developed with simulated $Gaia$ BP/RP spectra, with an ultimate objective of providing the community with both a recipe to select stars for medium/high resolution observations and a catalogue of stellar metallicities.}
   {We used a datased comprised of GALAH DR3 and SAGA database stars in order to verify and adjust to real world data our selection method. For that purpose, we used dereddening as a mean to tackle the issue of extinction, and then we applied our fine-tuned method to select metal-poor candidates, which we thereafter observed and analysed.}
   {We were able to infer metallicities for GALAH DR3 and SAGA stars - with color excesses up to $E(B-V)<1.5$ - with an uncertainty of $\sigma_{\mathrm{[Fe/H]}_{inf}} \sim 0.36$, which is good enough for the purpose of identifying new metal-poor stars. Further, we selected 26 metal-poor candidates - via our method - for observations. As spectral analysis showed, 100\% of them had $\mathrm{[Fe/H]}<-2.0$, 57\% had $\mathrm{[Fe/H]}<-2.5$ and 8\% had $\mathrm{[Fe/H]}<-3.0$. We inferred metallicities for these stars with an uncertainty $\sigma_{\mathrm{[Fe/H]}_{inf}}\sim0.31$, as was proven when comparing to the spectroscopic $\mathrm{[Fe/H]}$. Finally, we assembled a catalogue of metallicities for 10 861 062 stars.}
   {}
   {}

    \titlerunning{Metallicities for more than 10 million stars derived from $Gaia$ BP/RP spectra}
    \authorrunning{T.Xylakis-Dornbusch et al.}
   \maketitle
   
%

\section{Introduction}

The oldest stars that are still alive today and are located nearby, have metallicities $<-3$ (extremely metal-poor (EMP) stars \citep{2005ARA&A..43..531B}). EMP stars are rare and difficult to find. They are the descendants of the first generation of stars. Hence, EMP stars carry information that can shed light on the properties of their predecessors, as well as on how the latter exploded and ended their lives. Consequently, finding a large number of new EMP stars for which detailed studies of their chemical composition could be conducted is of the essence, since that would provide constraints on the assembly of the Galaxy, on the initial mass function of the first stars and on the nucleosynthesis processes that formed the heavy elements. The $Gaia$ Survey \citep{2016A&A...595A...1G} released in 2022 the low-resolution (R$\sim50$) $Gaia$ BP/RP spectra for 219 million sources \citep{2022arXiv220606143D}, and there have already been many studies that provide metallicity estimates for several thousands to millions of objects by extracting information from BP/RP spectra, often complementary to the use of additional data from $Gaia$ itself (for example Radial Velocity Spectrometer (RVS) spectra;  \citealt{2023A&A...674A...5K}) or other surveys. \cite{2023arXiv230410772B} derived metallicities for $\sim700\,000$ stars, and \cite{2022arXiv220606138A} delivered a catalog of stellar parameters - including the metallicity - using a Bayesian forward-modelling approach \citep{2013A&A...559A..74B}. \cite{2023arXiv230317676Y} used a classification algorithm (XGBoost; \citealt{Chen_2016}) to identify 188 000 very metal-poor star candidates. \cite{2022ApJ...941...45R} used the machine learning algorithm XGBoost to estimate $\mathrm{[M/H]}$ for 2 million stars, with 18 000 of them in the very- and metal-poor regime. \cite{2023ApJS..267....8A} produced a new catalogue, improving the \cite{2022ApJ...941...45R} one. The new catalogue was assembled by training the XGBoost algorithm on stellar parameters taken from the Data Release 17 (DR17) of the Sloan Digital Sky Survey's (SDSS) APOGEE survey \citep{2022ApJS..259...35A}, and from \cite{2022ApJ...931..147L} who derived stellar parameters for 400 extremely and ultra metal-poor stars . \cite{2023ApJS..267....8A} delivered a catalogue for $\sim$ 175 million stars, with a mean precision of 0.1 dex for $\mathrm{[M/H]}$. \cite{2023MNRAS.524.1855Z} used a forward model to estimate effective temperature, surface gravity, metallicity, distance, and extinction for 220 million stars. In order to do so, they used Gaia XP based data-driven models along with 2MASS \citep{2006AJ....131.1163S}, and WISE \citep{2019ApJS..240...30S} photometry. The forward model was then trained and validated on stellar parameters from the LAMOST survey \citep{2022ApJS..259...51W}, yielding $\mathrm{[Fe/H]}$ with a typical uncertainty of 0.15 dex. \cite{2023arXiv230801344M} used the BP/RP spectra to derive synthetic photometry of the Ca H \& K region, based on the narrow-band photometry of the Pristine Survey \citep{2017MNRAS.471.2587S}. They updated the Pristine metallicity inference model, so that it is exclusively based on Gaia magnitudes ($G$, $G_{BP}$, and $G_{RP}$), and produced a catalogue of metallicities for more than 52 million stars. \cite{2023arXiv230801344M} show that their photometric metallicities are accurate down to $\mathrm{[Fe/H]}\sim\,-3.5$, and thus very much suited for the study of the metal-poor Galaxy. Another study that take advantage of the BP/RP spectra in order to derive stellar parameters and/or metallicities is \cite{2023arXiv230708730C}. \newline 
\cite{2022A&A...666A..58X} (Paper I) developed an empirical method based on flux-ratios of synthetic $Gaia$ BP/RP spectra for the purpose of identifying new metal-poor stars. Specifically, the flux-ratios were those of the Ca H \& K lines to the H$\beta$ region ($fr\mathrm{_{CaHK/H\beta}}$ with $388\mathrm{nm}<\lambda<401\mathrm{nm}$ and $479\mathrm{nm}<\lambda<501\mathrm{nm}$), and the G-band region to the Ca near infrared triplet (Ca NIR) ($fr\mathrm{_{G/CaNIR}}$ with $420\mathrm{nm}<\lambda<444\mathrm{nm}$ and $846\mathrm{nm}<\lambda<870\mathrm{nm}$). It was shown that for roughly constant $fr\mathrm{_{G/CaNIR}}$, the $fr\mathrm{_{CaHK/H\beta}}$ is exponentially declining as metallicity increases. This work is a follow-up to Paper I, aiming at verifying the metal-poor star selection recipe presented therein by applying it to $Gaia$ DR3 BP/RP spectra. The paper is laid out in the following manner: in Section \ref{methods} we describe the dataset we used for the purpose of the validation of the method in Paper I, as well as how we addressed the issue of extinction which was not dealt with in our previous work. We close the Section with the description of the modifications we performed on the selection procedure and metallicity estimation of the metal-poor candidate stars compared to that introduced in Paper I. Next, we present in Section \ref{results} the results of the method verification, including the expected success rate in selecting stars that are very metal-poor ($\mathrm{[Fe/H]}<-2$) and below that threshold, and the purity of that ensemble. Then we investigate the plausibility of OBA stars being selected as metal-poor stars via our method. Furthermore, in Section \ref{selection} we describe the application of our fine-tuned recipe by selecting candidate metal-poor stars and subsequently observing them. We then present the results of our observations. Finally, in Section \ref{catalogue} we present a catalogue of metallicities including stars in both the metal-poor and -rich regimes.

\section{Methods}\label{methods}
For the verification of the selection process we used stellar parameters from high and medium resolution surveys/studies, and the respective flux dereddened Gaia BP/RP spectra. The software \code{GaiaXPy} \footnote{Software available at \href{https://gaia-dpci.github.io/GaiaXPy-website/}{https://gaia-dpci.github.io/GaiaXPy-website/}, version: DOI v2.0.1: 10.5281/zenodo.7566303} was used to generate the $Gaia$ BP/RP spectra, \code{dust\_extinction} \footnote{\href{https://github.com/karllark/dust\_extinction}{https://github.com/karllark/dust\_extinction}} and \code{dustmaps} \footnote{\href{https://github.com/gregreen/dustmaps}{https://github.com/gregreen/dustmaps}} \citep{2018JOSS....3..695M} were used to deredden the spectral flux.
 
 \subsection{Dataset}\label{dataset}
\begin{figure}
 \begin{minipage}{0.5\linewidth} 
   \centering
   \includegraphics[width=\textwidth]{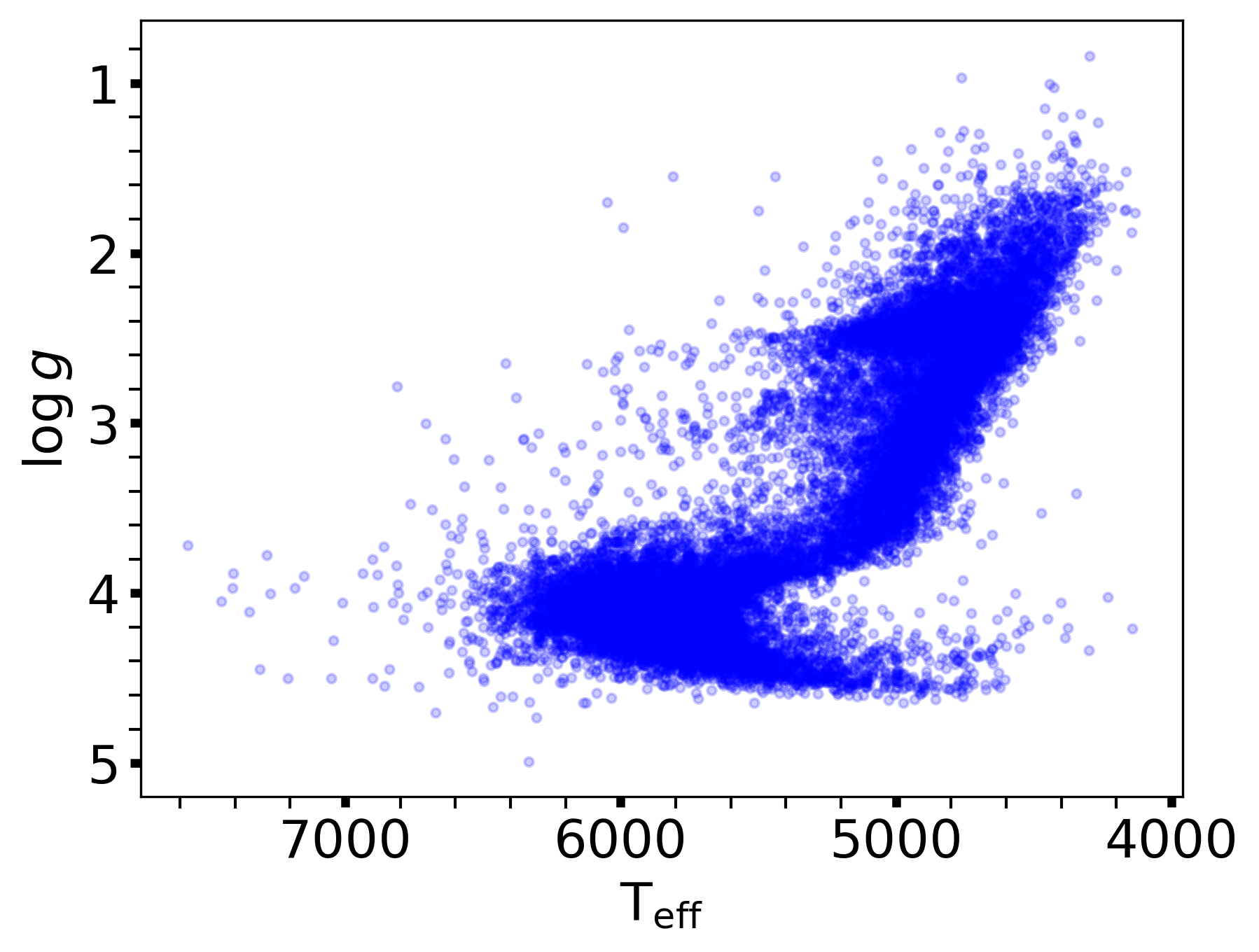}
   \caption{Kiel Diagram of the dataset.}
   \label{fig:kiel}
\end{minipage}%
 \begin{minipage}{0.5\linewidth}
   \centering
   \includegraphics[width=0.9\textwidth]{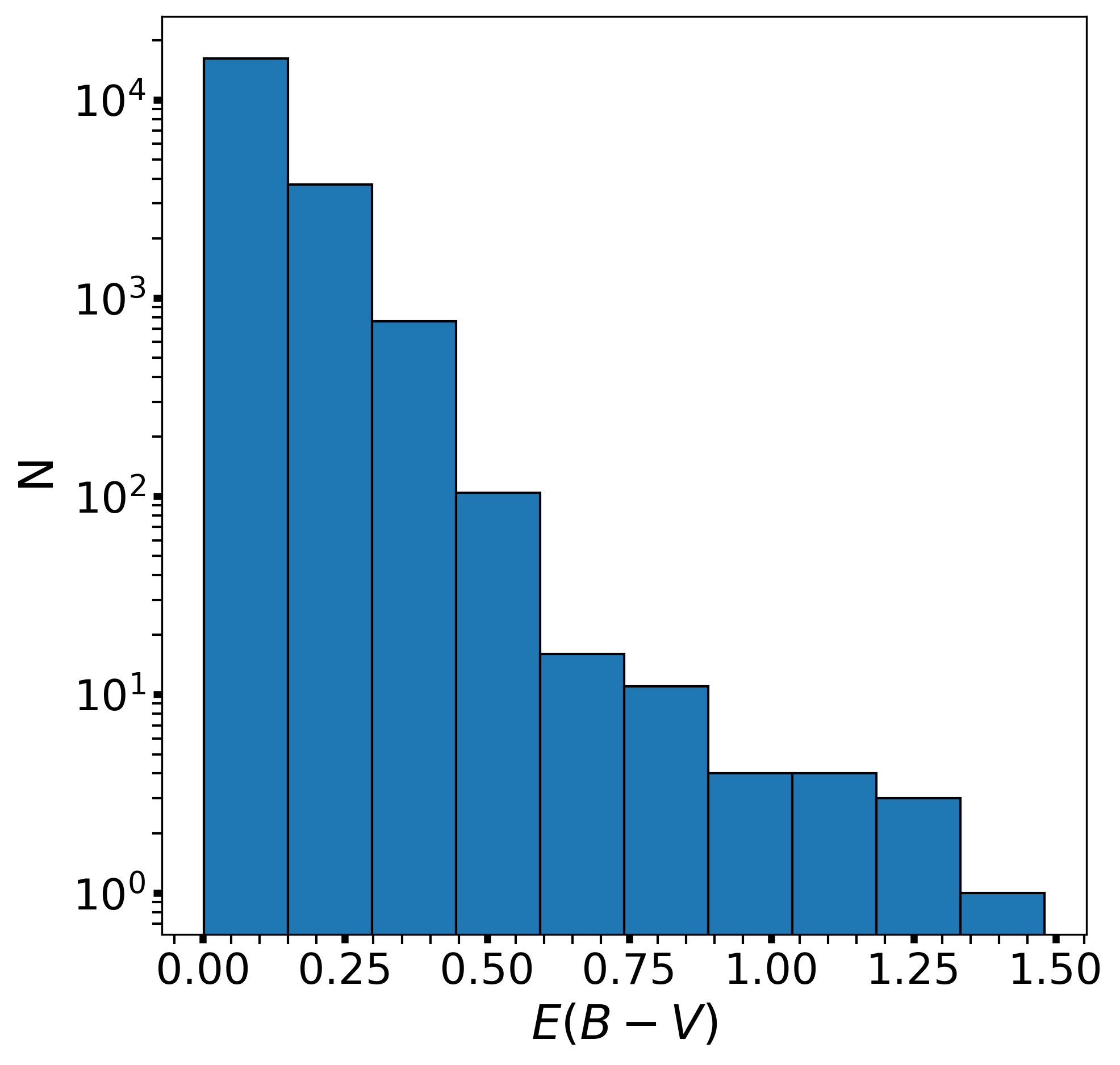}
   \caption{Histogram of the reddening distribution of our dataset.}
   \label{fig:dataset_hist}
\end{minipage}%
\end{figure}

 The dataset we used for this work is comprised of two different cross-matches with $Gaia$ BP/RP externally calibrated spectra \citep{2023A&A...674A...3M,2022arXiv220800211G,2016A&A...595A...1G}: the first one with the Stellar Abundances for Galactic Archaeology (SAGA) database \citep{2008PASJ...60.1159S,2011MNRAS.412..843S,2013MNRAS.436.1362Y,2017PASJ...69...76S}, and the second with the Galactic Archaeology with HERMES data release 3 (GALAH DR3) \citep{2021MNRAS.506..150B}. Both datasets together consist of 21 812 stars. We applied quality cuts on the aforementioned dataset by finding correlations between falsely identified metal-poor stars and quality parameters and ended up with 20 850 stars. Since this procedure could only be done after the application of our method to the dataset, we will be elaborating on it both in this section, as well as in the results section. The quality cuts we applied were twofold: one with respect to the quality of the stellar parameters of the dataset and another stemming from the quality of the $Gaia$ BP/RP spectra themselves, as well as of the effect of reddening. Concerning the first, stars for which there was no reliable metallicity estimate from GALAH were dropped (flag\_fe\_h=0). The mean uncertainty in the iron abundance for the remaining GALAH stars is 0.12 dex. We did not use any quality flag for the SAGA stars, but we resorted to the provided iron abundance uncertainties ($\overline{\sigma}_{SAGA} \sim 0.17$ dex). The GALAH $\mathrm{[Fe/H]}$ were computed using $\mathrm{A(Fe)_\odot=7.38}$ (for details see \citealt{2021MNRAS.506..150B}), while the SAGA database utilizes the \cite{asplund2009} solar chemical composition, that is, $\mathrm{A(Fe)_\odot=7.50}$. We consider this difference in the normalization of the metallicities of the two components comprising our dataset to be negligible, since we are not aiming at delivering high precision iron abundances, but rather intent to identify metal-poor stars. Further, as appears in the Kiel Diagram (Figure \ref{fig:kiel}), the final dataset we used spans from dwarf to giant stars, with most of the GALAH stars having disk-like kinematics \citep{2021MNRAS.506..150B} and a mean distance of $\overline{D}\sim1.9$ kpc (distances taken from \citealt{2018AJ....156...58B}), and the SAGA stars having $\overline{D}\sim 1.8$ kpc (distances taken from \citealt{2022arXiv220605992F}) and belonging to the Galactic halo. Regarding the spectra quality, we set a limit to the blending fraction $\beta$ of the BP/RP spectra and the color excess ($E(B-V)$). The former was defined by \cite{2021A&A...649A...3R} as ``...the sum of the number of blended transits in BP and RP divided by the sum of the number of observations in BP and RP''. We slightly modified it, specifically as 
 
 \begin{align*}
  \beta=\mathrm{(bp\_n\_blended\_transits+
rp\_n\_blended\_transits}+\\
\mathrm{bp\_n\_contaminated\_transits}+
\mathrm{rp\_n\_contaminated\_transits)}/\\
\mathrm{(bp\_n\_transits+rp\_n\_transits)}
 \end{align*}
  We set $\beta \leq 0.5$.

 Finally, complementary to our work in Paper I, we include objects in our dataset whose reddening is well above $E(B-V)=0.06$ (see Figure \ref{fig:dataset_hist}), which mandates that we tackle the issue of extinction.
 \subsection{Reddening}\label{redden}
 As a first approach we aimed at finding a reddening independent index, similar to \cite{2000A&AS..145..473B}. Since the region of the H$\beta$ line is part of the $fr_{\mathrm{CaHK/H\beta}}$ ratio (see Paper1 for details), we decided to test if the Str\"omgren $\beta$ index withstands extinction and replace the H$\beta$ region in $fr_{\mathrm{CaHK/H\beta}}$ with the former. The results were not those we anticipated: the $\beta$ index changed with extinction, even though it showed a sensitivity to effective temperature. As we were not able to define a reddening-independent metallicity calibration, we instead sought to implement reddening corrections for the metallicity calibration by means of dereddening the spectra. Therefore, we used the dust maps of \cite{1998ApJ...500..525S} (SFD) re-calibrated by \cite{2011ApJ...737..103S}, the extinction model of \cite{1999PASP..111...63F} and $R_v=3.1$ to deredden the externally calibrated BP/RP spectra. We repeated the above procedure using the extinction model of \cite{1989ApJ...345..245C} and found that the resulting flux-ratios have minimal differences with those calculated with the \cite{1999PASP..111...63F} model. We chose the SFD maps because they cover the entire sky. Considering the fact that the SFD maps account for the foreground dust, our stars need to be distant enough or at a galactic latitude great enough for the distance dependence to be neglected. The SAGA stars are halo stars, and thus distant enough (D $\geq1$ kpc ; \citealt{2011ApJ...737..103S}). In total, 81\% of the stars in our dataset are either at a distance D $\geq1$ kpc or at a latitude $\mid b \mid > 30\degree$. For the remaining 19\%, we calculated the reddening correction from \cite{2000AJ....120.2065B}. For most of the stars we found no or very small (<0.001 mag) correction. Only for 4\% of the total sample we found $E(B-V)$ corrections $\geq 0.02$ mag, so applying such a correction would have a negligible effect on the distribution in the dereddened flux-ratio plane (Figure \ref{fig:dataset_dered_AND_raw}).

 \subsection{Application of the method}
 In Paper I we provided coefficients for different pairs of $\mathrm{T_{eff}}$ and $\log g$ for the estimation of $\mathrm{[Fe/H]}$. We calibrated the coefficients for application to the real data, but the results were not corresponding to the theoretical expectations. Further, the problem of acquiring well-defined effective temperatures and surface gravities for millions of stars -so that the metal-poor ones among them could be identified- became apparent. We decided to use only quantities that can be directly derived from the spectra, i.e. the flux-ratios. The plane of the $fr_{\mathrm{CaHK/H\beta}}$ and $fr_{\mathrm{G/CaNIR}}$ flux-ratios (see Figure \ref{fig:dataset_dered_AND_raw}), enables us to find the loci of metal-poor ($\mathrm{[Fe/H]}<-1.0$) and further metal-deficient stars. The grey lines in Figure \ref{fig:dataset_dered_AND_raw} represent different metallicity regimes, with the stars below the dashed-dotted and dotted ones being metal-poor ($\mathrm{[Fe/H]}<-1$) and very metal-poor ($\mathrm{[Fe/H]}<-2$) respectively.
 
\section{Results}\label{results}

\begin{figure}
    \centering \includegraphics[width=0.5\textwidth]{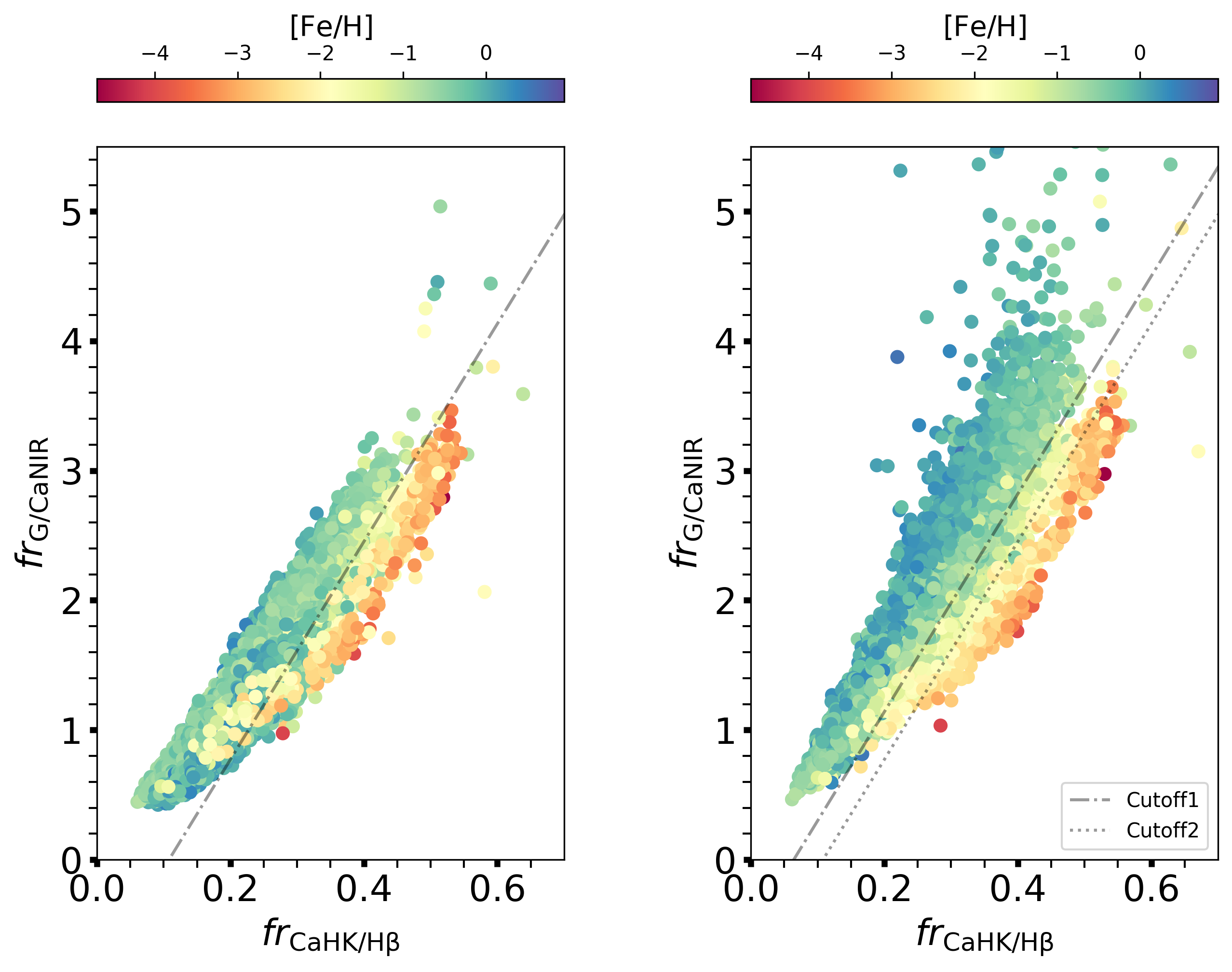}
    \caption{Flux-ratios of raw (left) and dereddened (right) fluxes from $Gaia$ BP/RP spectra are shown in the left and right panel respectively. The color-coding reflects the metallicity of the stars of the dataset we used. Below the dashed-dotted and dotted lines are the flux-ratio areas where stars with $\mathrm{[Fe/H]}\leq-1$ and $\mathrm{[Fe/H]}\leq-2$ respectively, are primarily found.}
    \label{fig:dataset_dered_AND_raw}
\end{figure}

\begin{figure}
 \begin{minipage}{0.5\linewidth} 
   \centering
   \includegraphics[width=\textwidth]{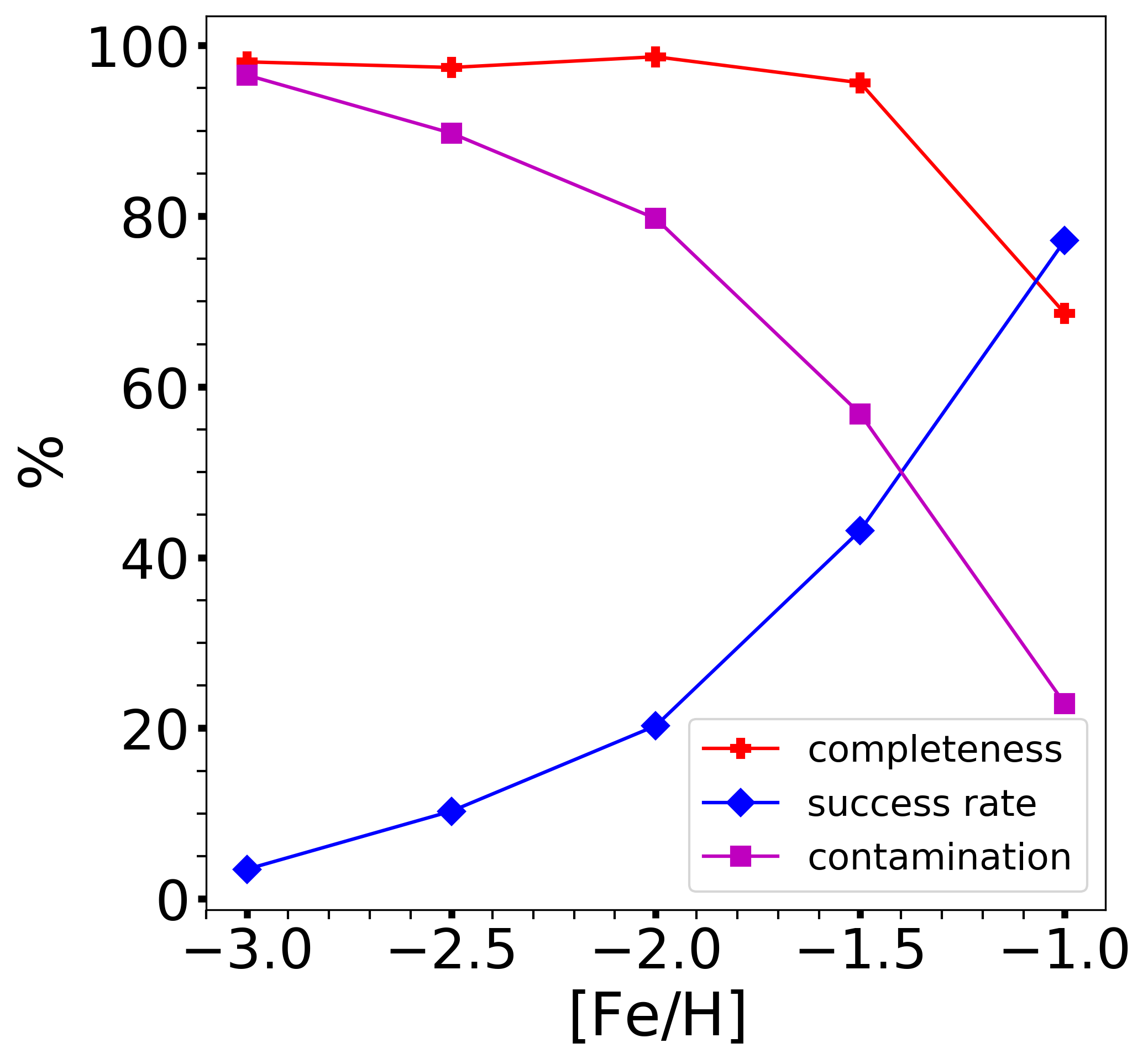}
   \end{minipage}%
   \centering
    \begin{minipage}{0.5\linewidth} 
    \includegraphics[width=\textwidth]{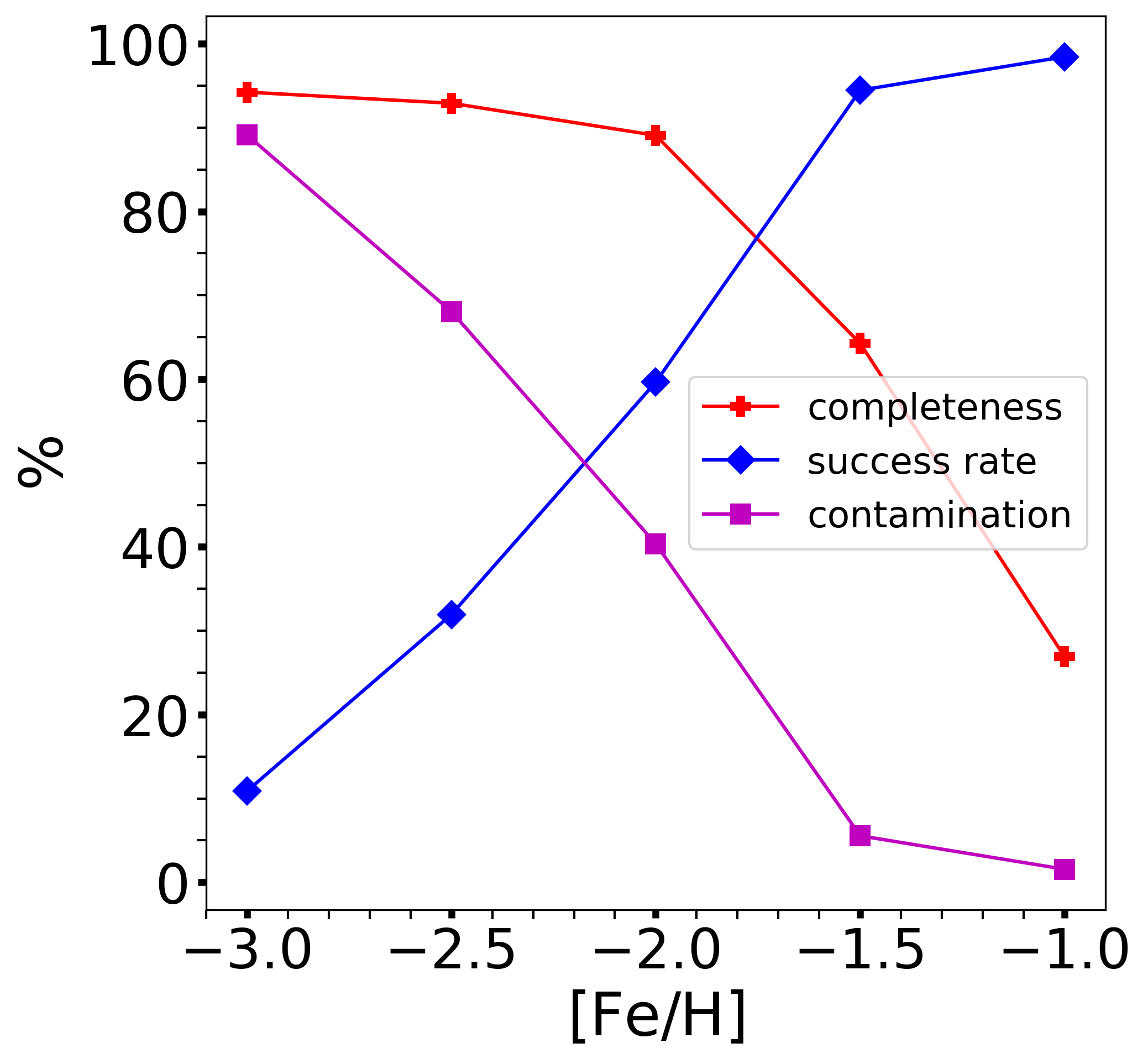}
    \end{minipage}
    \caption{Completeness, success rate and contamination of the stars that were selected from below the dashed-dotted (left panel) and dotted line (right panel). The stars were selected from a dereddened flux-ratio plane.}
   \label{fig:statistics}
\end{figure}

The results in the right panel of Figure \ref{fig:dataset_dered_AND_raw}, depict a clear correlation between metallicity, $fr_{\mathrm{CaHK/H\beta}}$ and $fr_{\mathrm{G/CaNIR}}$ flux-ratios. The left panel shows the flux-ratios before dereddening and the right panel shows the dereddened values. We overplot a dashed-dotted (Cutoff1) and a dotted line (Cutoff2), to designate flux-ratio areas where objects with $\mathrm{[Fe/H]}_{ref}\leq-1$ and $\mathrm{[Fe/H]}_{ref}\leq-2$ respectively, are primarily found ($\mathrm{[Fe/H]}_{ref}$ is the reference $\mathrm{[Fe/H]}$). By selecting metal-poor stars in that way, we found that there was a correlation between a high blending fraction $\beta$ and contaminants, i.e. stars with $\mathrm{[Fe/H]}_{ref}>-1$. We chose the $\beta$ such that there is a balance between acceptable contamination and completeness, since a greater $\beta$ means a greater number of stars. We define the completeness as the ratio of the number of selected stars below a certain metallicity threshold to the total number of stars in the dataset that have $\mathrm{[Fe/H]}_{ref}\leq \mathrm{threshold}$, the success rate as the percent of the selected stars that have $\mathrm{[Fe/H]}_{ref}$ below a certain specified value, and the contamination as the percent of selected stars that have a metallicity above the specified threshold.\newline
The results in Figure \ref{fig:dataset_dered_AND_raw} were generated after the application of the quality cuts described above. By choosing all the stars below Cutoff1 in Figure \ref{fig:dataset_dered_AND_raw}, we are able to recover from the GALAH-SAGA sample more than 98\% of the stars with  $\mathrm{[Fe/H]}\leq-2$, all the ultra metal-poor stars ($\mathrm{[Fe/H]}\leq-4$) and ~70\% of stars $\mathrm{[Fe/H]}<-1$. We record a success rate of $\sim 80\%$, 44\%  and 20\% for stars with $\mathrm{[Fe/H]}\leq-1,-1.5$ and -2 respectively. When we select stars below Cutoff2 we make a trade-off between the success rate and the completeness: we still recover more than ~90\% and ~94\% of very and extremely metal-poor stars respectively, but lose about 40\% of those with $-2<\mathrm{[Fe/H]}\leq-1$ compared to the other cut off. The success rate increases significantly to $\sim 99\%$, 95\%  and 60\% for stars with $\mathrm{[Fe/H]}\leq-1,-1.5$ and -2 respectively (summarized in Figure \ref{fig:statistics}). For comparison, we selected all stars that fall below the same -as before- dotted and dashed-dotted line without dereddening (left panel of Figure \ref{fig:dataset_dered_AND_raw}) and calculated the statistics as above. Even though the completeness for different metallicity categories are fairly similar and in some cases even slightly better, the success rate is much lower and in consequence the contamination is much higher. 

\begin{figure}
    \centering \includegraphics[width=0.35\textwidth]{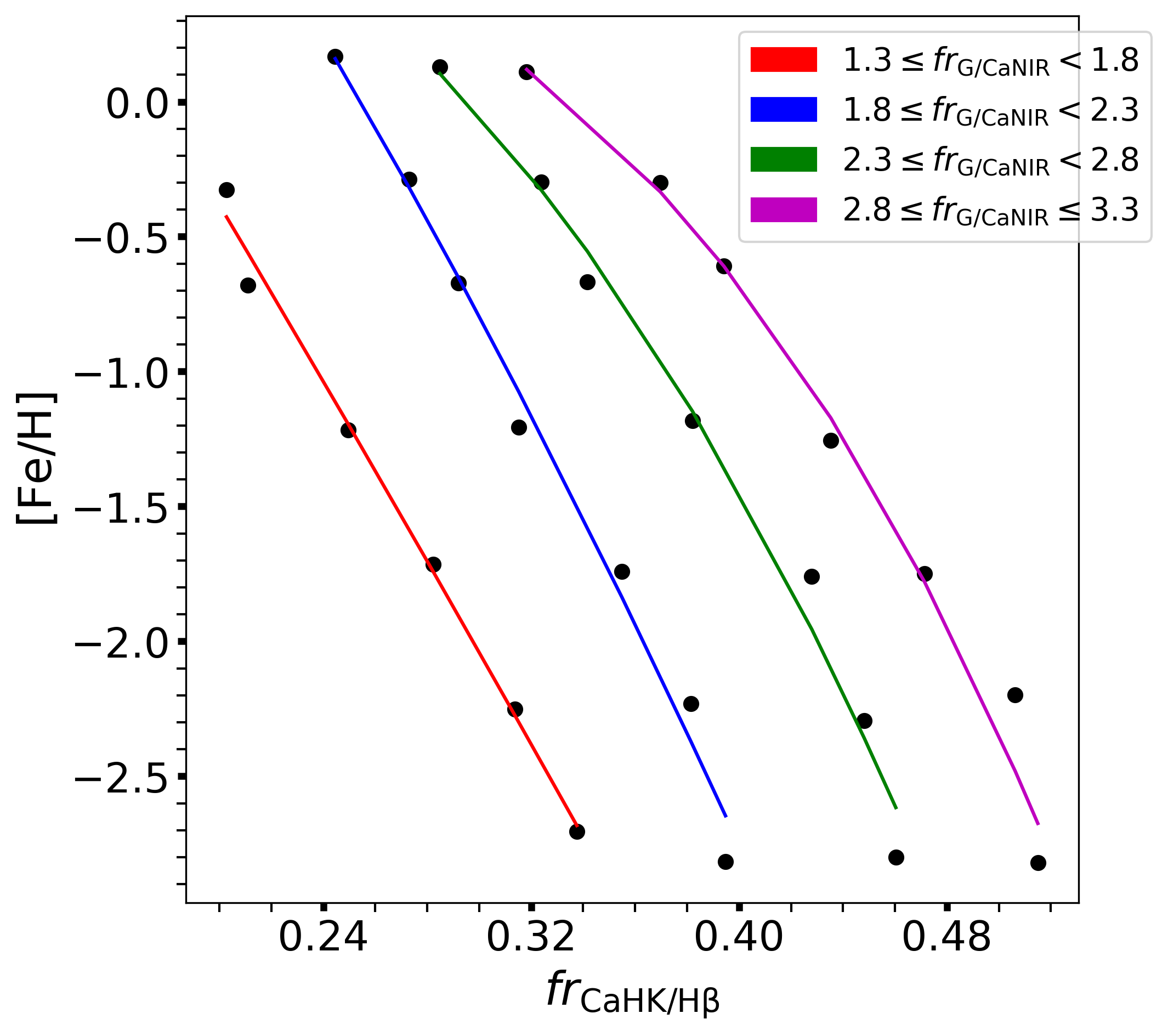}
    \caption{Best fits to the $fr_{\mathrm{CaHK/H\beta}}$ - $\mathrm{[Fe/H]}$ pairs. The different line colors convey the $fr\mathrm{_{G/CaNIR}}$ range of applicability.}
    \label{fig:feh_fit}
\end{figure}

\begin{figure}
    \centering \includegraphics[width=0.4\textwidth]{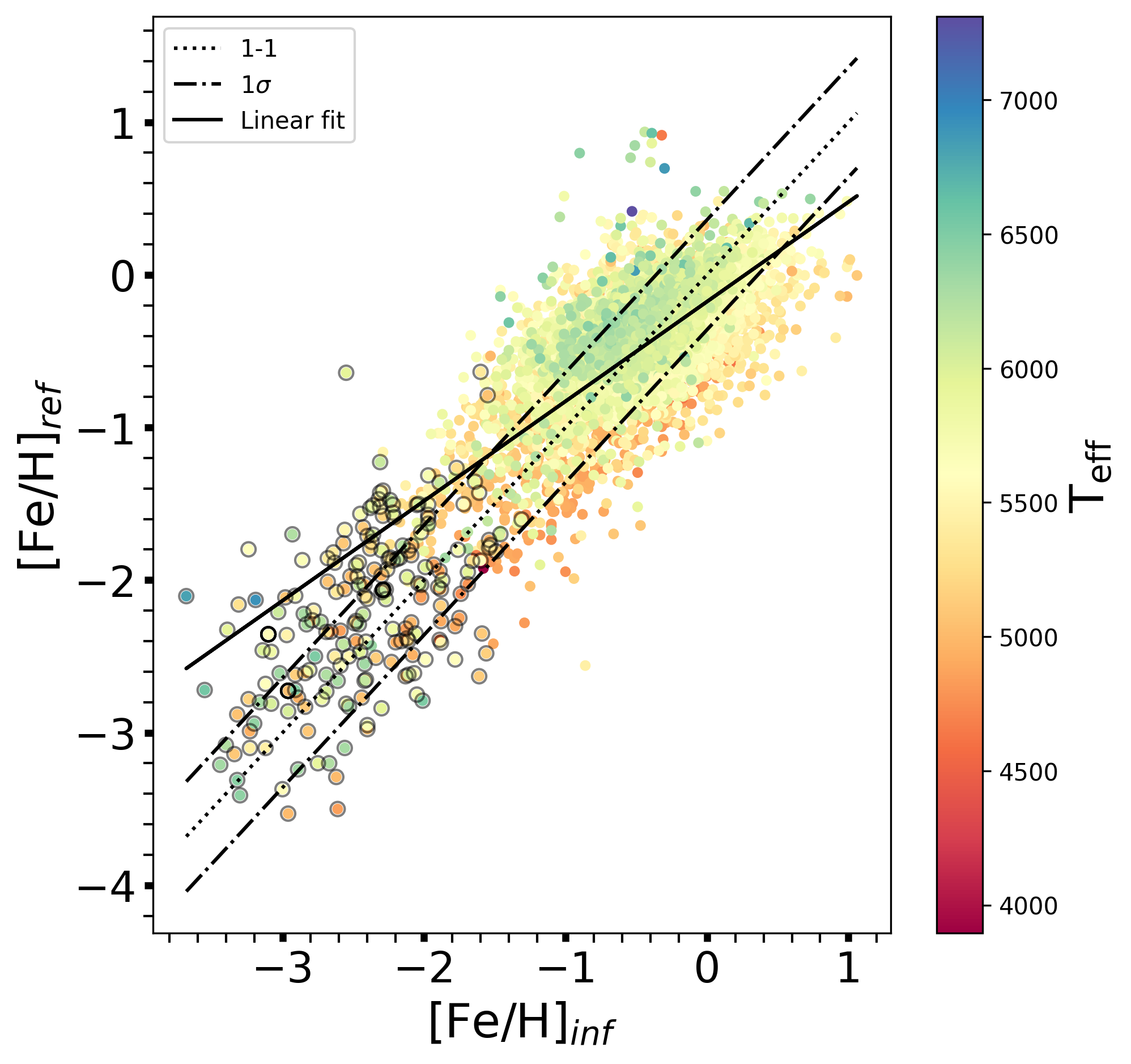}
    \caption{Metallicity estimation of a subset of the GALAH-SAGA dataset. The points that have a black circle around them, are located below the black-dotted line in the flux-ratio plane (Figure \ref{fig:dataset_dered_AND_raw}). The color-coding reflects the effective temperature of the stars. We plot the inferred and reference $\mathrm{[Fe/H]}$ on the x- and y-axis respectively. }
    \label{fig:feh_estimation}
\end{figure}

Further, we find that by selecting the metal-poor candidates through the flux-ratio plane, we can extrapolate the theoretical method described in Paper I, to a broader parameter space. Specifically, in Paper I the recipe was developed for FGK stars in the effective temperature range of 4800-6300 K, and in this study we retrieve metal-poor stars that have $4636\,K\leq\mathrm{T_{eff}}\leq7150\,K$.\newline
Finally, we estimated the iron abundances of our dataset as follows. First, we randomly sampled our GALAH-SAGA dataset, and split it into two equal parts. We divided the flux-ratios of the first sampled sub-dataset into $fr\mathrm{_{G/CaNIR}}$ bins. Then, we split each of those bins into metallicity bins, for which we calculated the mean $fr_{\mathrm{CaHK/H\beta}}$. Next we found best fits to the sets of $fr_{\mathrm{CaHK/H\beta}}$ - $\mathrm{[Fe/H]}$ pairs (Figure \ref{fig:feh_fit}), which we subsequently used to estimate the iron abundance of the second sub-dataset. We used the following function in order to perform the fittings:
\begin{equation}
   fr\mathrm{_{G/CaNIR}} = -a\cdot fr\mathrm{_{CaHK/H\beta}}^b+c\label{eq:1}
\end{equation}
where $a,b$ and $c$ are the coefficients of the best fit, which are shown in Table \ref{table:coefs}. The respective results of the metallicity estimation are presented in Figure \ref{fig:feh_estimation}. We were able to infer $\mathrm{[Fe/H]}$ with an uncertainty of $\sigma_{\mathrm{[Fe/H]}_{inf}} \sim 0.36$ dex. This precision is sufficient for reliably identifying metal-poor stars.

\begin{table}
\caption{Coefficients of the best fit.}             
\label{table:coefs}      
\centering                          
\begin{tabular}{c c c c}        
\hline\hline                 
$a$ & $b$ & $c$ & $fr\mathrm{_{G/CaNIR}}$ \\    
\hline                        
   17.497139 & 1.119316 &2.506009 & [1.3-1.8) \\
22.219935 & 1.512732 & 2.800237 & [1.8-2.3) \\
29.101554 & 2.624439 & 1.18252 & [2.3-2.8) \\
32.268827 & 3.351201 & 0.815609 & [2.8-3.3] \\
\hline                                   
\end{tabular}
\tablefoot{The $fr\mathrm{_{G/CaNIR}}$ values, are the ranges of applicability of each set of coefficients.}
\end{table}

\section{OBA stars}\label{OBA_stars}
OBA stars are young hot stars that can present emission lines in their spectra. When OB stars are highly reddened, they can appear as K-type stars, hence good reddening values are essential in order to tell them apart from metal-poor FGK stars. Also, young or accreting stars can show emission lines at various spectral regions, including the Ca H\&K absorption lines. Consequently, the emission in the Ca II H\&K lines results in a net weak absorption line, masking those stars as being metal-poor. Therefore we wish to test to which degree those stars are expected to contaminate a selected metal-poor-candidate sample.

\begin{figure}
    \centering \includegraphics[width=0.4\textwidth]{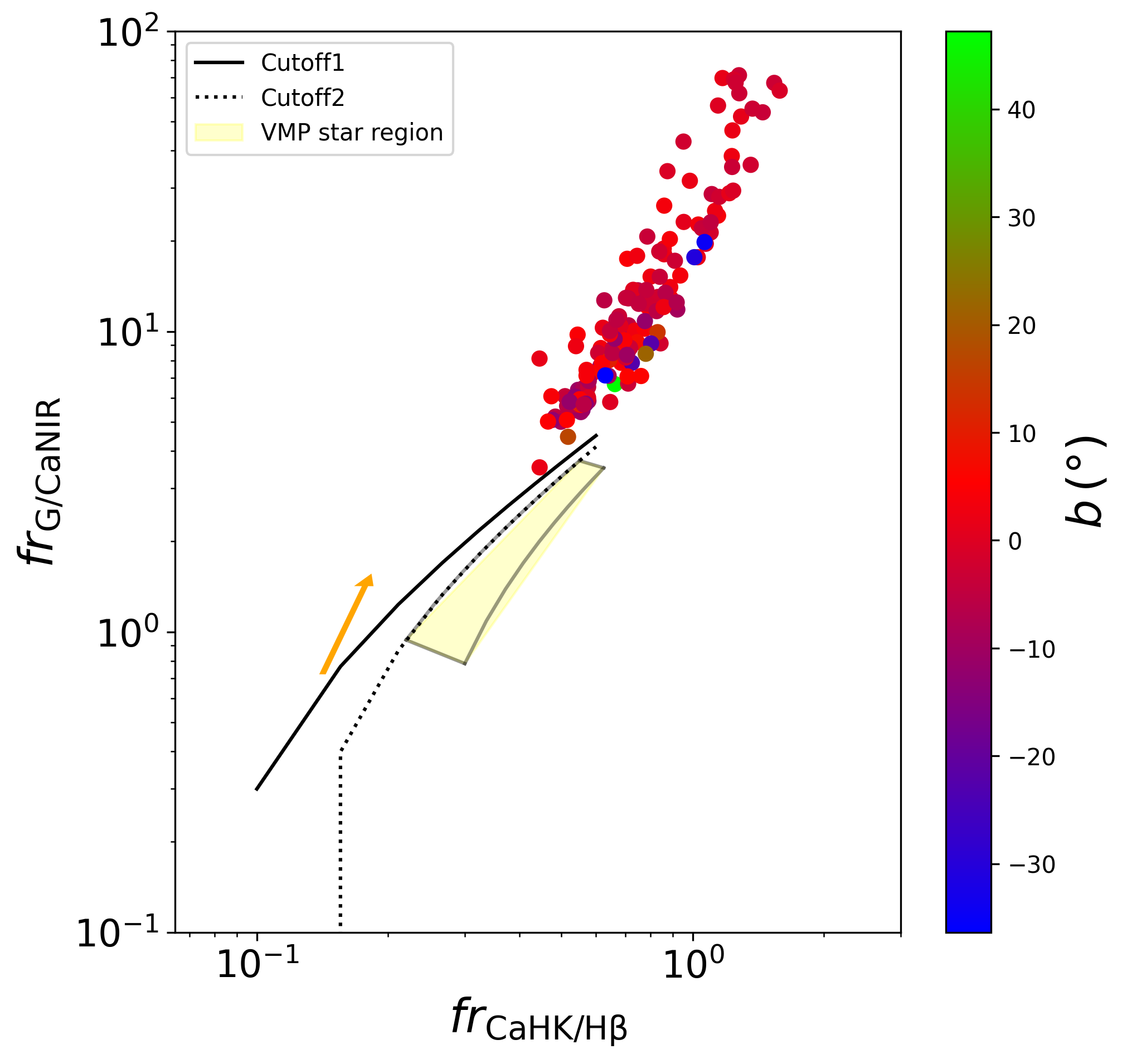}
    \caption{Flux-ratios of OBA stars. The solid and dotted lines represent the Cutoff1 and Cutoff2 respectively, while the yellow shaded area designates the region that is populated by very-metal poor (VMP) stars (see Figure \ref{fig:dataset_dered_AND_raw}). The color-coding indicates the Galactic Latitude $b$ of each star. As can be seen, most of the stars are located on the Galactic plane ($\mid b\mid\leq10 \degree$),. The orange arrow illustrates the effect of extinction for a color excess $E(B-V)\approx0.3$ mag. The nock and the point of the arrow represent the flux-ratios before and after dereddening, respectively.}
    \label{fig:OBA_stars}
\end{figure}

We select a random subset of 200 stars from the OBA stars golden sample \citep{https://doi.org/10.17876/gaia/dr.3/59}. From those, 193 stars have an externally calibrated BP/RP spectrum, and 173 have a blending fraction $\beta \leq 0.5$. We deredden the externally calibrated spectra as described in section \ref{redden} and subsequently compute the flux-ratios. In Figure \ref{fig:OBA_stars} we plot the flux-ratios of the OBA golden sample subset. In order to show the effect of extinction, which depends on the color-excess rather than on the flux-ratios, we use logarithmic axes. The effect of extinction is demonstrated with an arrow (orange arrow), where its nock and point represent the flux-ratios before and after dereddening, respectively, for $E(B-V)\approx0.3$ mag.  As can be seen, none of the 173 stars appear in the region of the flux-ratio plane where the metal-poor stars frequent (Figure \ref{fig:dataset_dered_AND_raw}). However, due to the fact that the location of the stars on the flux-ratio plane depends on the extinction, we caution the reader that highly reddened OBA stars with underestimated color-excess values could appear in the region of metal-poor stars (yellow area in Figure \ref{fig:OBA_stars}) and hence, contaminate the sample of metal-poor stars selected via this method. 

\section{Observational metal-poor star candidate verification}\label{selection}

In order to verify our metal-poor candidate selection method as well as the metallicity estimation presented herein, we selected a sample of stars from $Gaia$ DR3, which had not been observed before. We opted to select fairly bright giant stars, in order to achieve a good enough signal-to-noise ratio (SNR) for the purpose of deriving precise $\mathrm{[Fe/H]}$. Further, the location of the telescope to be used was known beforehand, hence we used the following selection criteria: G=12-13 mag, Ra=16-02h, Dec= 00$\degree$-+20$\degree$, $\mid b \mid>20\degree$ and $\beta \leq 0.5$, which rendered 90 798 stars. Following, we computed the flux-ratios. From the 90 798 stars, we chose those with flux-ratios $1\leq fr\mathrm{_{G/CaNIR}}\leq5$, which left us with 70 509 stars. Then, we selected all the stars below a more stringent cut than Cutoff2, which is a line that is shifted parallel to Cutoff2 by 0.1+$fr\mathrm{_{CaHK/H\beta}}$. That cutoff left us with 77 stars, of which 10 stars had already been observed in high resolution and their metallicities are -or will be- in the literature. It is worth noting, that all the 10 stars that appear in literature, are metal-poor. The reason we used a more stringent cut was, that there is a clear correlation between the inferred metallicity and the position of the star on the flux-ratio plane. We opted to observe candidates with the lowest predicted metallicities, so if we would have used Cutoff2, most of the stars above the more stringent cut off would have not made it into the final target list due to the higher estimated $\mathrm{[Fe/H]}_{inf}$. We show the distribution of the inferred $\mathrm{[Fe/H]}_{inf}$ for metal-poor candidates that were located between Cuttoff2 and our chosen cut off in Figure \ref{fig:above_cut}. Finally, we estimated the $\mathrm{[Fe/H]}$ for the remaining 67 stars, and our final target list was comprised of 32 stars with $\mathrm{[Fe/H]}_{inf}\leq-2.35$, of which we managed to observed 26. Of the 35 stars that were not included in the target list, 8 of them were outside the metallicity inference range ($fr\mathrm{_{G/CaNIR}}>3.3$). The distribution of the inferred metallicities for the remaining 27 metal-poor candidates that were not included in the final target list is show in Figure \ref{fig:not_included}.

\subsection{Observations and metallicity determinations}\label{observations}

The targets were observed at the McDonald Observatory with the Harlan J. Smith 2.7m telescope and the TS23 echelle spectrograph \citep{tull1995}. The spectra were obtained using a 1.2" slit and 1x1 binning, yielding a resolving power of $R\sim 60,000$ and covering a wavelength range of 3600-10000 \AA. The 26 stars were observed over four nights in August 2023. The data was reduced using standard IRAF packages \citep{tody1986,tody1993}, including correction for bias, flatfield, and scattered light. Table \ref{table:observations} lists the Gaia DR3 id, right ascension, declination, Heliocentric Julian Date (HJD), exposure times, the signal-to-noise ratio per pixel (SNR) at 5000$\AA$ and heliocentric radial velocities. The heliocentric radial velocities were determined via cross-correlation with a spectrum of the standard star HD~182488 ($V_{hel}=-21.2$ kms$^{-1}$; \citet{gaiarv}) obtained on the same run.

We determined stellar parameters ($\mathrm{T_{eff}}$, $\log g$, $\mathrm{[Fe/H]}$, and $v_t$) for the observed stars from a combination of photometry and equivalent width (EW) measurements of \ion{Fe}{i} and \ion{Fe}{ii} lines, and using the software \code{smhr}\footnote{\href{https://github.com/andycasey/smhr}{https://github.com/andycasey/smhr}} \citep{casey2014} to run the radiative transfer code \code{MOOG}\footnote{\href{https://github.com/alexji/moog17scat}{https://github.com/alexji/moog17scat}} \citep{sneden1973,sobeck2011} assuming local thermodynamical equilibrium. We used one dimensional plane-parallel, $\alpha$-enhanced ($\mathrm{[\alpha/Fe] = +0.4}$) stellar model atmospheres computed from the \code{ATLAS9} grid \citep{castelli2003}, line lists from \code{linemake}\footnote{\href{https://github.com/vmplacco/linemake}{https://github.com/vmplacco/linemake}} \citep{placco2021}, and Solar abundances were taken from \citet{asplund2009}.  $\mathrm{T_{eff}}$ for the stars was determined from dereddened $Gaia$ $G$, $BP$, $RP$ \citep{anders2022,babusiaux2018} and 2MASS $K$ magnitudes \citep{cutri2003} using the color-$\mathrm{T_{eff}}$ relations from \cite{mucciarelli2021}. For the $K$ magnitudes, we used the extinction coefficient from \citet{mccall2004}. The $\log g$ were then determined by requiring ionisation equilibrium between the \ion{Fe}{i} and \ion{Fe}{ii} lines and $v_t$ by 
requiring no correlation of \ion{Fe}{i} line abundances with reduced EW. Final $\mathrm{[Fe/H]_{spec}}$ of stars is taken as the mean abundances of the \ion{Fe}{i} lines and the uncertainties are the standard deviation of these. Final stellar parameters are listed in Table \ref{table:observations}.

\subsection{Results}\label{mp_candid}
\begin{figure}
    \centering \includegraphics[width=0.4\textwidth]{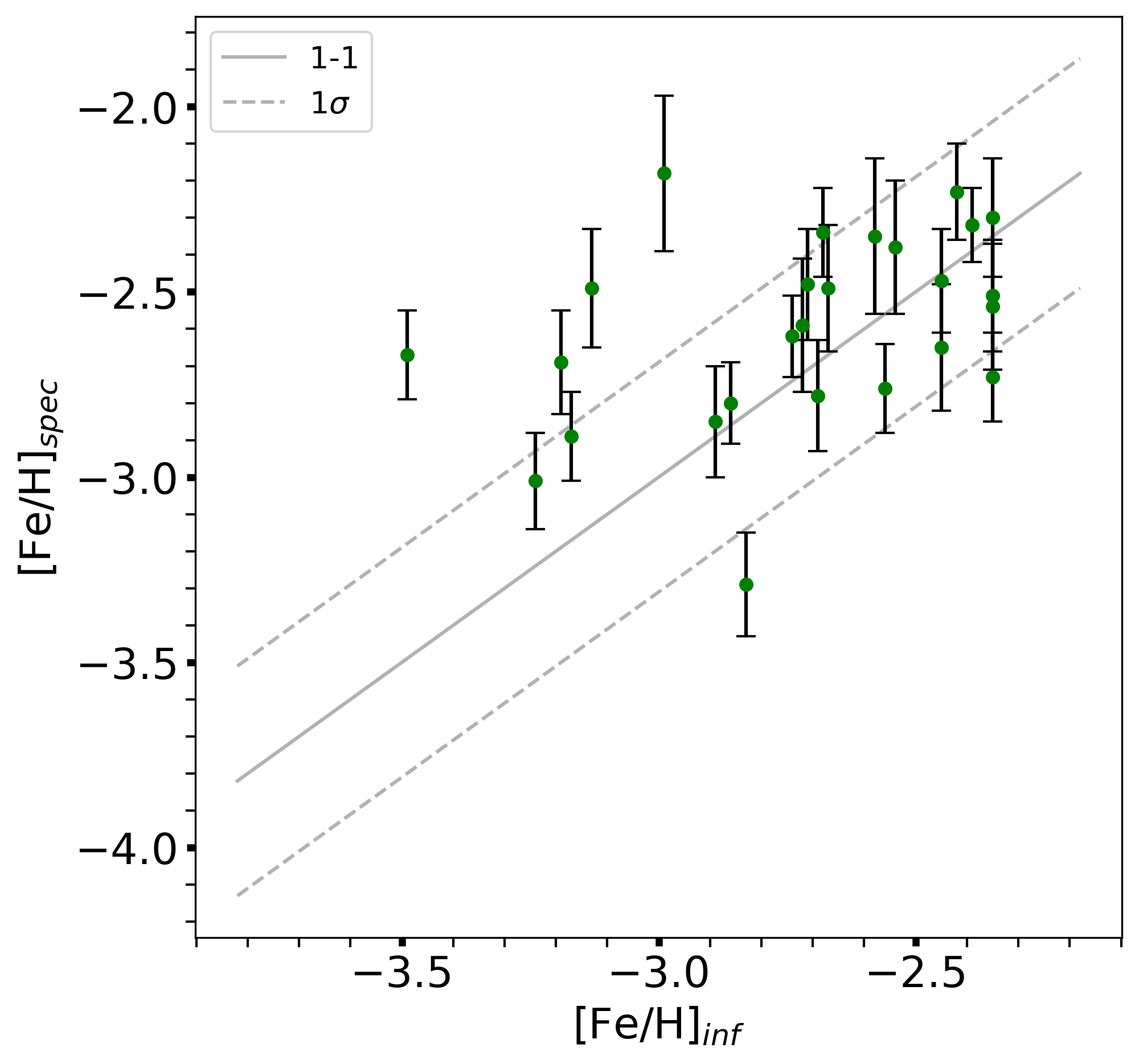}
    \caption{$\mathrm{[Fe/H]}_{inf}$ versus $\mathrm{[Fe/H]}_{spec}$. The solid grey line is the 1 to 1 line, and the dashed grey line designates the 1$\sigma$ uncertainty ($\sigma_{\mathrm{[Fe/H]}_{spec}}=0.31$  dex) in $\mathrm{[Fe/H]}_{spec}$.}
    \label{fig:feh_inf_vs_spec}
\end{figure}

The stellar parameters of the observed stars are shown in Table \ref{table:observations}. As the parameters show, all observed stars are metal-poor FGK stars. The uncertainty in our metallicity inference is $\sigma_{\mathrm{[Fe/H]}_{inf}}\sim 0.31$, which agrees with the uncertainty in deriving metallicities for the GALAH-SAGA sample ($\sigma_{\mathrm{[Fe/H]}_{inf}}\sim 0.36$), as described above. Figure \ref{fig:feh_inf_vs_spec} shows $\mathrm{[Fe/H]}_{inf}$ versus the spectroscopic determined $\mathrm{[Fe/H]_{spec}}$. Further, 100\% of the observed stars are very metal-poor, 58\% have $\mathrm{[Fe/H]}<-2.5$ and 8\% are EMP. Lastly, we did not have any contamination from OBA stars, which agrees with our finding in Section \ref{OBA_stars}.

\begin{table*}[h]
        \caption{Stellar Parameters and observation log of observed metal-poor candidates.}
        \label{table:observations}
    \centering
    \resizebox{\textwidth}{!}{%
    \begin{tabular} {c c c c r c r c c c c c c c c}
    \hline\hline
    Gaia DR3 ID & RA (J2000) & Dec (J2000) & HJD & exp time & SNR & $V_{hel}\pm\sigma$ &$\mathrm{[Fe/H]}_{inf}$ & $G$ & $\mathrm{T_{eff}}$ & $\log g$ & $\mathrm{[Fe/H]}_{spec}$  & $\sigma_{\mathrm{[Fe/H]}_{spec}}$ & $v_t$ & $E(B-V)$\\ 
     & (hrs) & ($\degree$) & & (s) & \@5000\AA & (kms$^{-1}$) &(dex) & (mag) & (K) & (dex) & (dex) & (dex) & (kms$^{-1}$) & (mag)\\\hline
    4560234719702983552 & 17 00 58.86 &	+18 12 48.76 & 2460177.61 & 3x1200 & 38 &$-200.2\pm0.7$ & -3.49 & 12.95 & 5512 & 1.81 & -2.67 & 0.12 & 2.26 & 0.067 \\
    2770306858573498880 & 23 54 36.90 &	+13 43 33.61 & 2460175.83 & 3x900  & 43 & $-91.8\pm0.9$ & -3.19 & 12.75 & 5280 & 2.42 & -2.69 & 0.14 & 1.63 & 0.043\\
    2740778202499153280 & 00 09 13.91 &	+03 34 27.64 & 2460175.88 & 3x1200 & 29 & $-63.3\pm0.7$ & -3.17 & 12.52 & 5172 & 2.18 & -2.89 & 0.12 & 1.58 & 0.024\\
    2783063972298129280 & 00 42 31.57 & +18 34 52.58 & 2460175.93 & 3x1200 & 23 &$-349.7\pm0.7$ & -3.13 & 12.53 & 5463 & 1.92 & -2.49 & 0.16 & 1.98 & 0.058\\
    38721161695303808   & 03 55 46.04 &	+13 28 40.99 & 2460178.96 & 3x1200 & 11 &  $59.1\pm1.1$ & -2.99 & 12.89 & 5666 & 3.09 & -2.18 & 0.21 & 1.95 & 0.298\\
    1788649988097920768 & 21 20 25.52 &	+19 16 40.19 & 2460175.75 & 3x1200 & 35 &   $2.1\pm1.4$ & -3.24 & 12.84 & 5868 & 1.95 & -3.01 & 0.13 & 2.65 & 0.078\\
    2739719922558093440 & 23 50 58.88 & +02 36 12.99 & 2460176.84 & 3x600  & 38 &  $15.4\pm1.3$ & -2.89 & 12.10 & 6488 & 3.40 & -2.85 & 0.15 & 1.77 & 0.033\\
    3268830653286376704 & 03 15 35.79 & +02 25 49.29 & 2460176.96 & 3x1200 & 27 & $203.3\pm0.7$ & -2.72 & 12.81 & 5114 & 2.18 & -2.59 & 0.18 & 1.49 & 0.095\\
    4446252678577892224 & 16 34 16.18 &	+08 49 40.19 & 2460178.63 & 3x900  & 22 & $-12.1\pm0.8$ & -2.71 & 12.27 & 4995 & 1.88 & -2.48 & 0.15 & 1.81 & 0.065\\
    2719036833232602752 & 22 57 03.19 &	+12 58 25.60 & 2460175.80 & 3x600  & 41 &$-242.8\pm0.5$ & -2.68 & 12.15 & 5316 & 1.90 & -2.34 & 0.12 & 2.06 & 0.047\\
    4229999872631438848 & 20 32 11.41 &	+01 02 05.16 & 2460177.71 & 3x1200 & 37 &$-240.6\pm0.5$ & -2.67 & 12.82 & 5221 & 2.71 & -2.49 & 0.17 & 1.36 & 0.097\\
    1757147197551005952 & 21 09 25.50 &	+11 48 44.80 & 2460175.70 & 3x1200 & 26 &  $71.0\pm1.3$ & -2.86 & 12.60 & 5440 & 1.23 & -2.80 & 0.11 & 2.29 & 0.108\\
    1730672812979631104 & 20 56 02.56 &	+02 07 13.56 & 2460178.69 & 3x1200 & 14 &$-160.8\pm1.0$ & -2.58 & 12.84 & 5194 & 1.09 & -2.35 & 0.21 & 2.38 & 0.107\\
    2814304091236720000 & 23 27 36.00 &	+15 23 54.70 & 2460177.80 & 3x1200 & 31 & $-81.5\pm0.9$ & -2.56 & 12.96 & 5096 & 1.95 & -2.76 & 0.12 & 2.05 & 0.054\\
    4561199025759521920 & 17 02 37.48 &	+19 17 21.35 & 2460177.66 & 3x1200 & 31 &$-174.7\pm0.5$ & -2.54 & 12.79 & 5267 & 2.57 & -2.38 & 0.18 & 1.73 & 0.088\\
    4503007613380083328 & 17 53 57.89 &	+17 45 19.99 & 2460176.64 & 3x1200 & 50 &$-100.9\pm1.0$ & -2.69 & 12.42 & 6030 & 4.22 & -2.78 & 0.15 & 1.72 & 0.081\\
    4449403019908847488 & 16 48 40.46 &	+13 32 43.83 & 2460176.61 & 3x600  & 36 & $195.9\pm0.7$ & -2.45 & 12.15 & 5214 & 2.33 & -2.65 & 0.17 & 1.48 & 0.060\\
    2574400790177777408 & 01 58 30.71 &	+11 18 42.38 & 2460177.93 & 3x1200 & 38 &$-101.6\pm1.0$ & -2.42 & 12.58 & 6547 & 4.04 & -2.23 & 0.13 & 1.62 & 0.116\\
    2554217295745049856 & 00 39 13.34 &	+04 23 33.16 & 2460176.87 & 3x1200 & 38 &$-145.0\pm0.9$ & -2.39 & 12.81 & 6235 & 3.51 & -2.32 & 0.10 & 1.55 & 0.027\\
    2580053787477560576 & 01 19 10.86 &	+10 07 08.75 & 2460177.85 & 3x1200 & 40 & $-29.0\pm0.6$ & -2.45 & 12.55 & 5139 & 2.33 & -2.47 & 0.14 & 1.68 & 0.070\\
    2732958716319826048 & 22 33 41.08 &	+14 49 05.58 & 2460178.73 & 3x900  & 20 &  $26.8\pm0.9$ & -2.35 & 12.23 & 5636 & 2.59 & -2.54 & 0.17 & 1.24 & 0.069\\
    2756350516963035904 & 23 40 19.64 &	+05 34 00.66 & 2460176.80 & 3x1200 & 33 & $111.7\pm1.7$ & -2.35 & 12.91 & 6235 & 2.40 & -2.73 & 0.12 & 1.50 & 0.081\\
    2698131578134995456 & 21 34 27.89 &	+04 04 38.23 & 2460177.76 & 3x900  & 39 &$-299.1\pm0.6$ & -2.35 & 12.27 & 5294 & 2.90 & -2.51 & 0.15 & 1.39 & 0.054\\
    4432234794379005952 & 16 34 54.04 &	+02 06 14.94 & 2460178.60 & 3x900  & 17 &  $85.5\pm0.8$ & -2.35 & 12.23 & 5257 & 1.60 & -2.30 & 0.16 & 1.99 & 0.062\\
    2706127364131151744 & 22 41 26.08 &	+05 07 30.91 & 2460178.77 & 3x1000 & 21 &$-139.7\pm2.1$ & -2.83 & 12.65 & 5648 & 2.56 & -3.29 & 0.14 & 2.09 & 0.077\\
    1733398605383859840 & 21 09 34.21 &	+05 14 05.85 & 2460176.69 & 3x1200 & 30 & $-52.3\pm0.7$ & -2.74 & 12.70 & 5200 & 1.26 & -2.62 & 0.11 & 1.97 & 0.112\\

\hline
\end{tabular}%
}
\end{table*}

\section{Catalogue of stellar $\mathrm{[Fe/H]}$}\label{catalogue}

For the purpose of providing the community with a catalogue of metallicities, we used the following criteria from ''The Milky Way Halo High-Resolution Survey´´ \citep{2019Msngr.175...26C} of the 4-metre Multi-Object Spectroscopic Telescope ($4MOST$) \citep{https://doi.org/10.18727/0722-6691/5117}, combined with those developed for this work, to select stars from $Gaia$ DR3: 
\begin{itemize}
    \item $\mid b \mid>10\degree$ 
    \item $0.15\,\mathrm{mag}\leq\mathrm{(BP-RP)}_{0}<1.1\,$mag
    \item blending index $\beta \leq 0.5$
    \item $1.3\leq fr\mathrm{_{G/CaNIR}}\leq3.3$
    \item $E(B-V)\leq1.5\,$mag
\end{itemize}
The above criteria turned over 10 861 062 stars, for which we estimated the metallicity. 225 498 stars in this catalogue have $\mathrm{[Fe/H]}_{inf}<-2.0$. Further, in our catalogue 2236 stars have $\mathrm{[Fe/H]}_{inf}<-5.0$, which suggests that these stars have probably emission lines rather than being metal-poor. We cross-matched the stars of our catalogue that have $\mathrm{[Fe/H]}_{inf}<-2.0$ with the Gaia OBA golden sample \citep{https://doi.org/10.17876/gaia/dr.3/59}, and found that 104 of those stars are indeed OBA stars. Out of those OBA contaminants, 8 have an estimated metallicity $\mathrm{[Fe/H]}_{inf}<-5.0$ in the catalogue. Finally, a sample of the catalogue is shown in Table \ref{table:catalogue}.
\begin{table*}[h]
\caption{Sample of the metallicities catalogue.}             
\label{table:catalogue}
\centering
\resizebox{\textwidth}{!}{%
\begin{tabular}{c c c c c  c c c c c}
\hline\hline
$source\_id$ & $RA (J2000)$ & $DEC (J2000)$ & $E(B-V)$ & $fr\mathrm{_{CaHK/H\beta}}$ & $fr\mathrm{_{G/CaNIR}}$ & $\mathrm{[Fe/H]}_{inf}$ & $G$ & $G_{BP}$ & $G_{RP}$ \\
 & (\degree) & (\degree) & (mag) &  &  & (dex) & (mag) & (mag) & (mag) \\
\hline
1736084918450522624 & 311.672374 & 6.454326 & 0.086352 & 0.395785 & 3.040589 & -0.65 & 14.839189 & 15.116240 & 14.387665 \\
1736086121041383936 & 311.640473 & 6.501470 & 0.080705 & 0.386366 & 2.951276 & -0.53 & 12.410907 & 12.699553 & 11.959683 \\
1736086464640617472 & 311.605184 & 6.517305 & 0.079810 & 0.384081 & 3.279787 & -0.51 & 12.523436 & 12.753575 & 12.055906 \\
1736086769579627008 & 311.686028 & 6.547942 & 0.081594 & 0.440737 & 2.951817 & -1.27 & 14.591786 & 14.878093 & 14.132823 \\
1736089213417835264 & 311.740409 & 6.639877 & 0.088562 & 0.383000 & 2.963972 & -0.50 & 14.826837 & 15.120335 & 14.363322 \\
1736089934974194688 & 311.614227 & 6.572804 & 0.078512 & 0.320748 & 2.815581 & 0.09 & 12.647016 & 12.943399 & 12.197193 \\
1736090381648971776 & 311.524228 & 6.550632 & 0.083777 & 0.382586 & 2.909278 & -0.49 & 13.581572 & 13.875311 & 13.124882 \\
1736093847685000960 & 311.868064 & 6.611197 & 0.092579 & 0.410637 & 3.239939 & -0.84 & 14.073834 & 14.347256 & 13.636151 \\
1736099693138064384 & 311.706790 & 6.756967 & 0.085753 & 0.351283 & 2.938921 & -0.17 & 13.712016 & 14.000586 & 13.259321 \\
1736099693138065408 & 311.698148 & 6.754874 & 0.084601 & 0.376363 & 2.852484 & -0.42 & 15.257664 & 15.554148 & 14.791958 \\
\hline                                   
\end{tabular}
}
\tablefoot{The color excess values, $E(B-V)$, are taken from \cite{1998ApJ...500..525S} (re-calibrated by \cite{2011ApJ...737..103S}). The full catalogue is available at the CDS.}
\end{table*}

\section{Comparison to other catalogues}\label{cat_comparison}

\begin{figure*}
 \begin{minipage}{0.29\linewidth} 
   \centering
   \includegraphics[width=0.9\textwidth]{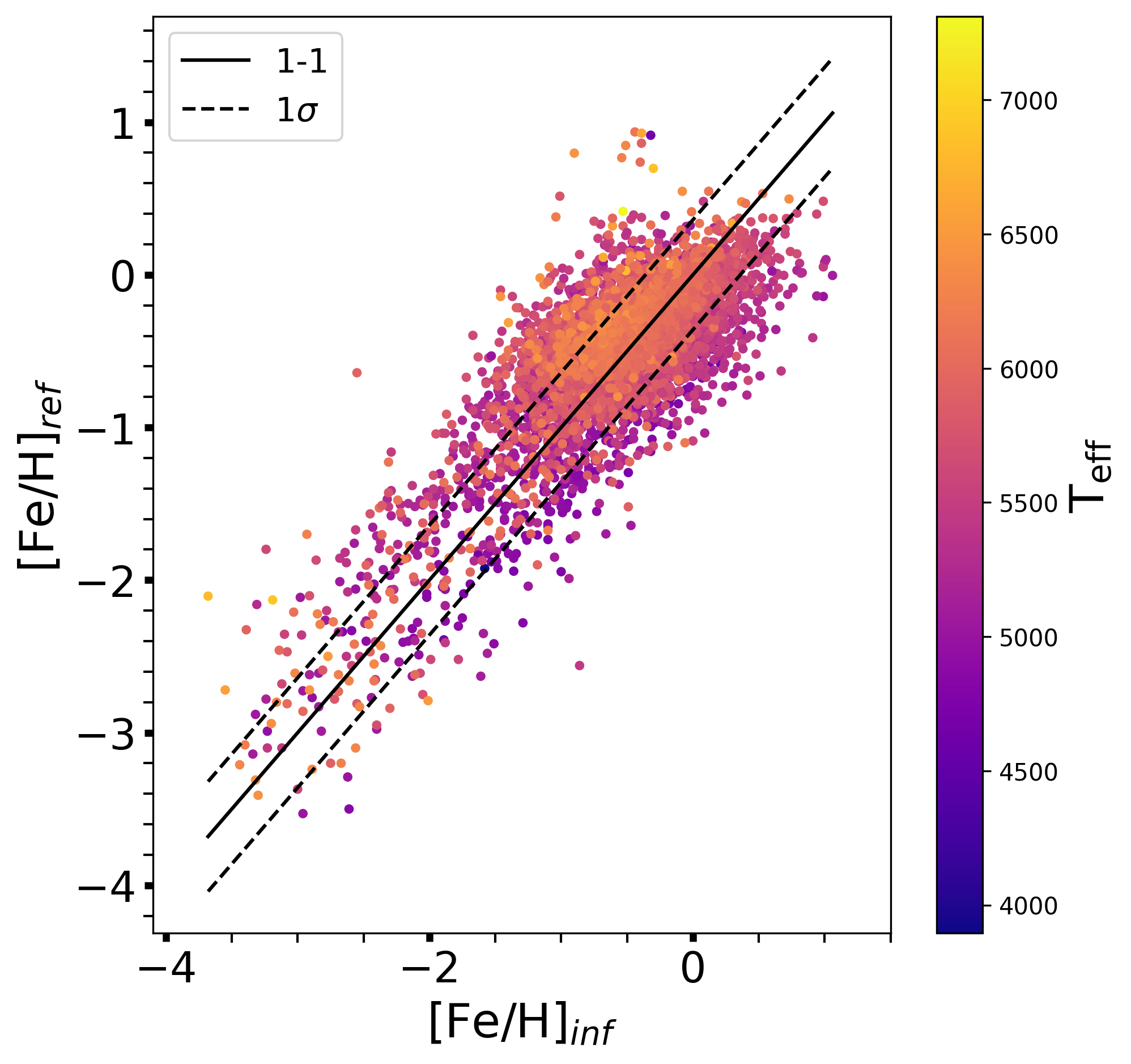}
   \end{minipage}%
   \centering
    \begin{minipage}{0.29\linewidth} 
    \includegraphics[width=0.9\textwidth]{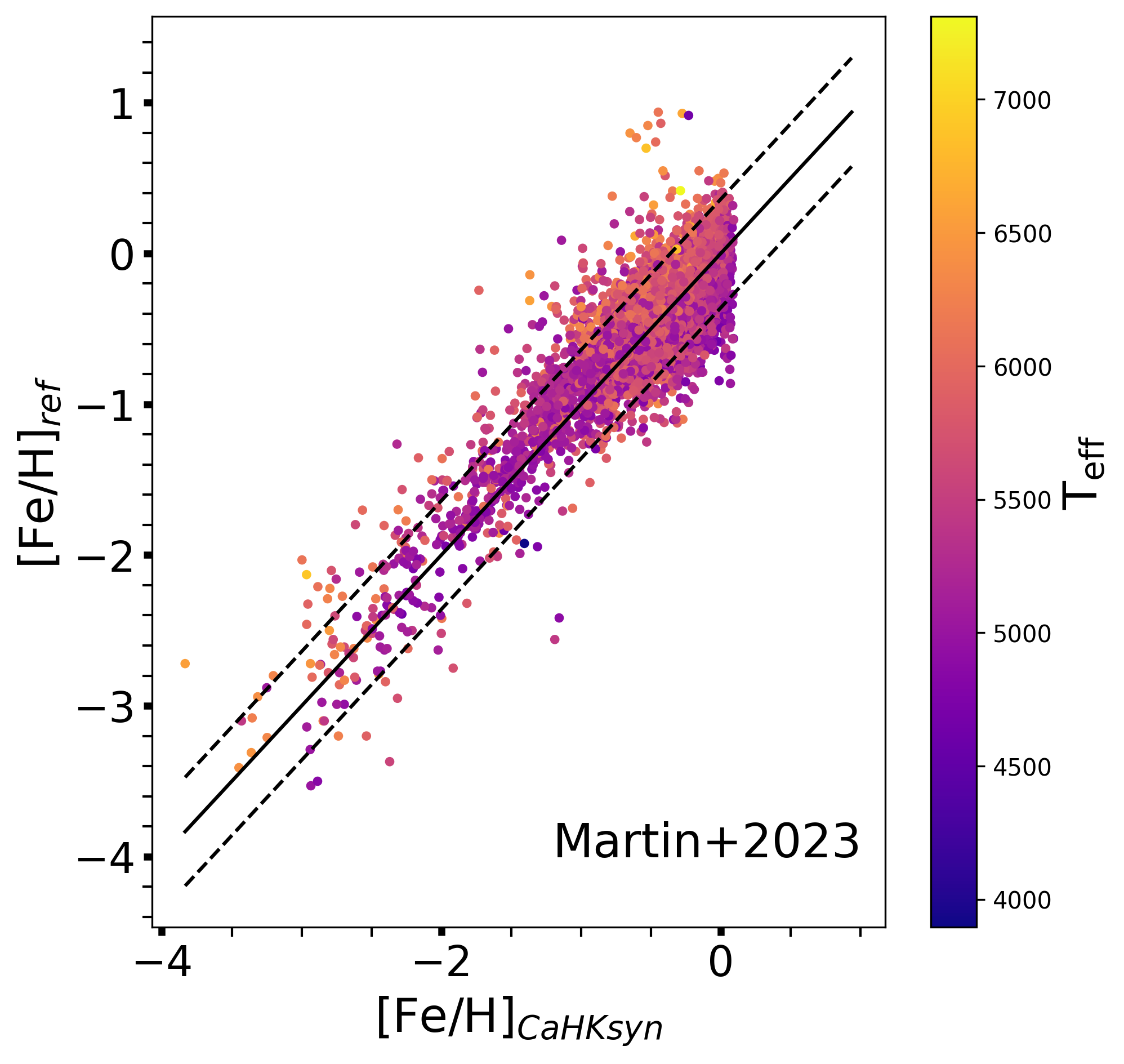}
    \end{minipage}
    \centering
    \begin{minipage}{0.29\linewidth} 
    \includegraphics[width=0.9\textwidth]{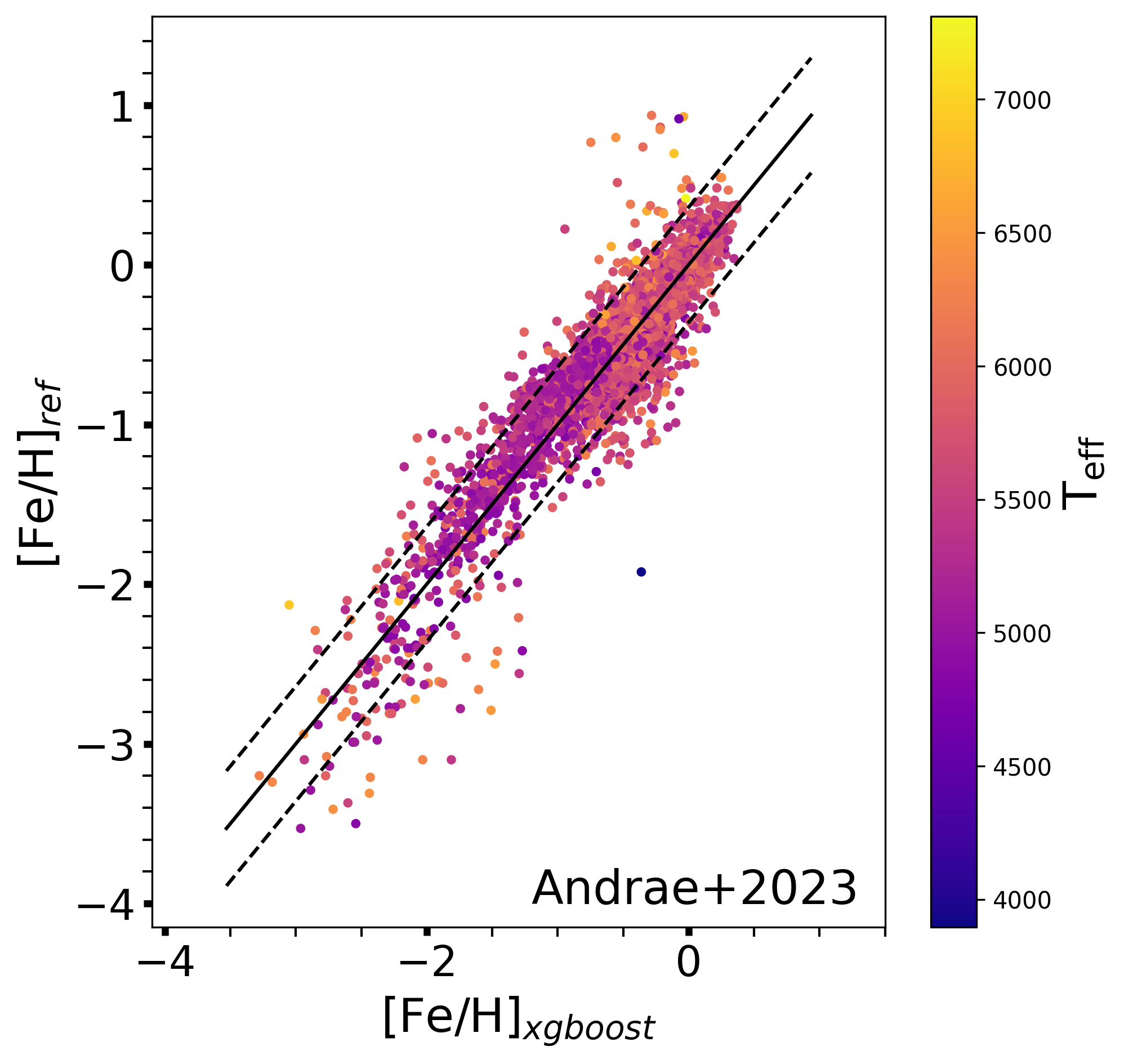}
    \end{minipage}
    \begin{minipage}{0.29\linewidth} 
   \centering
   \includegraphics[width=0.9\textwidth]{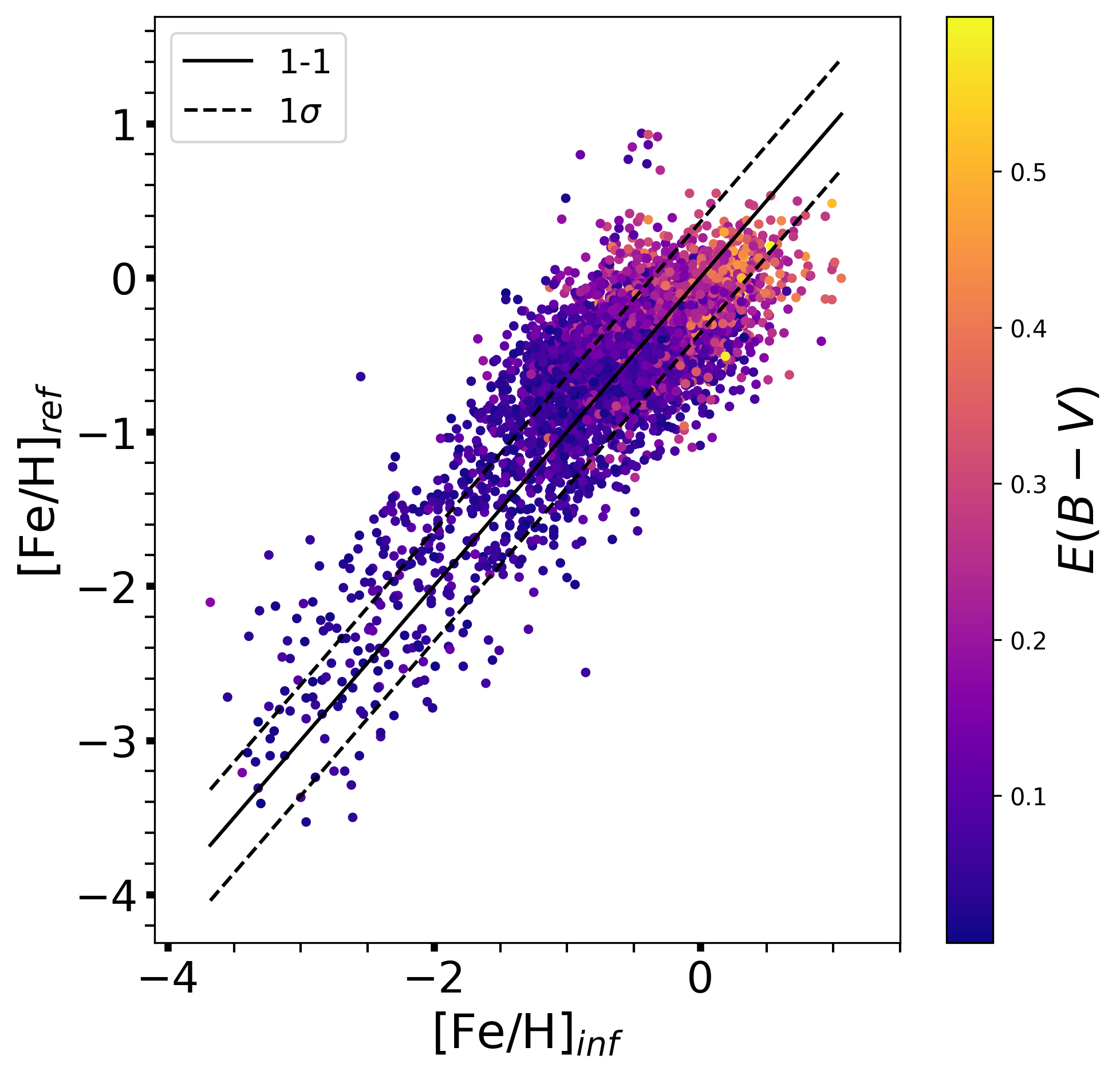}
   \end{minipage}%
   \centering
    \begin{minipage}{0.29\linewidth} 
    \includegraphics[width=0.9\textwidth]{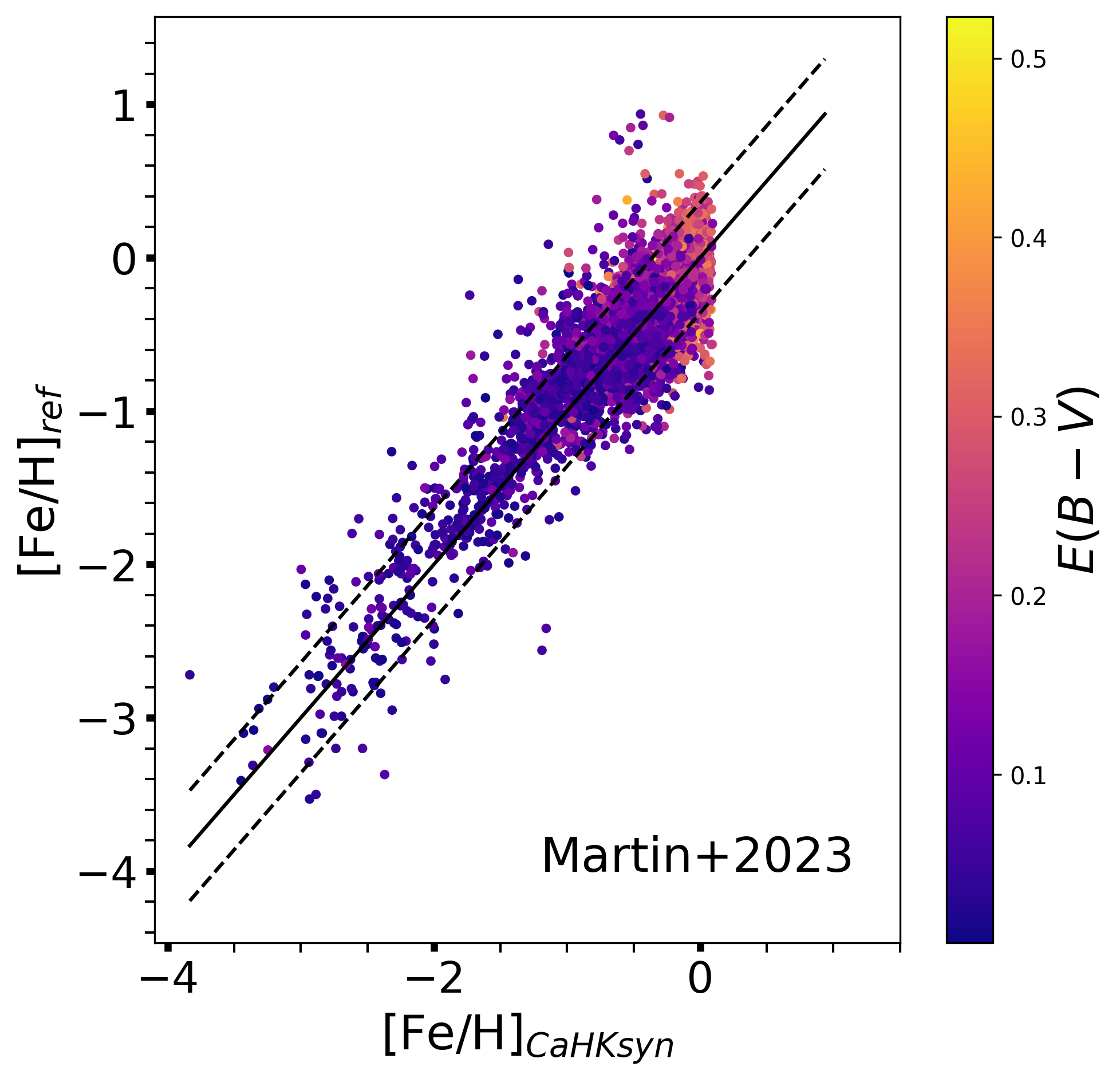}
    \end{minipage}
    \centering
    \begin{minipage}{0.29\linewidth} 
    \includegraphics[width=0.9\textwidth]{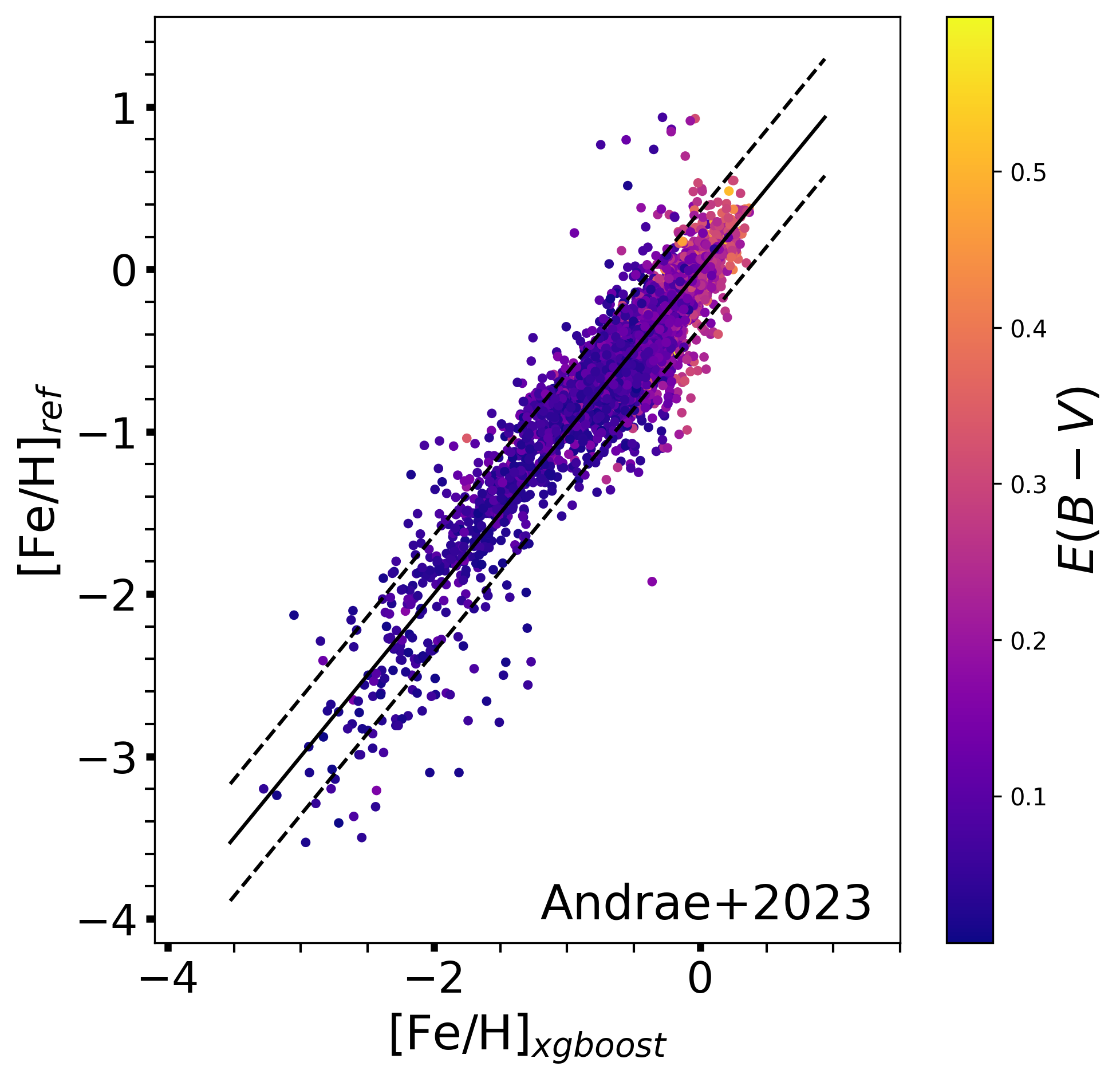}
    \end{minipage}
    \begin{minipage}{0.29\linewidth} 
   \centering
   \includegraphics[width=0.9\textwidth]{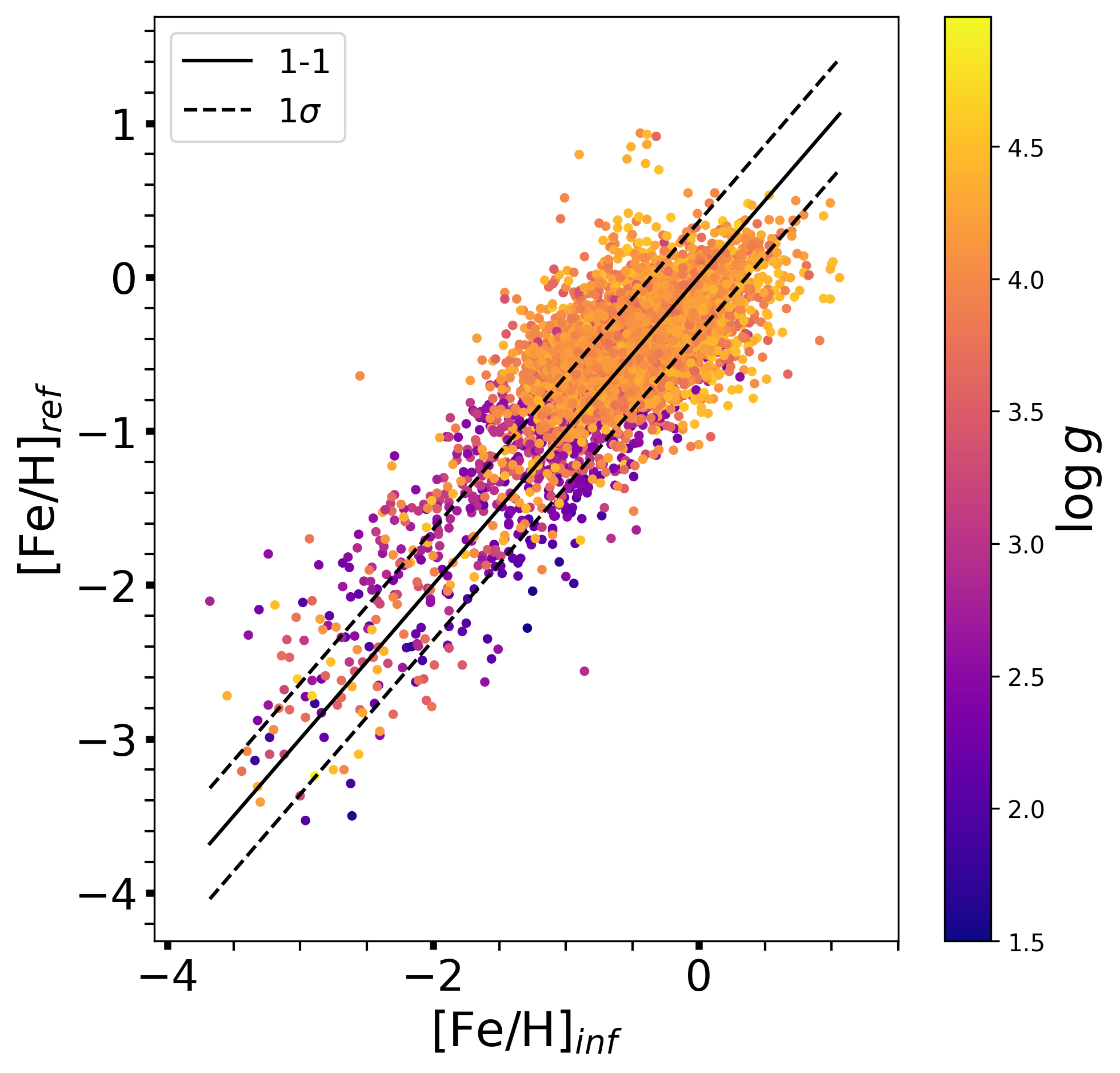}
   \end{minipage}%
   \centering
    \begin{minipage}{0.29\linewidth} 
    \includegraphics[width=0.9\textwidth]{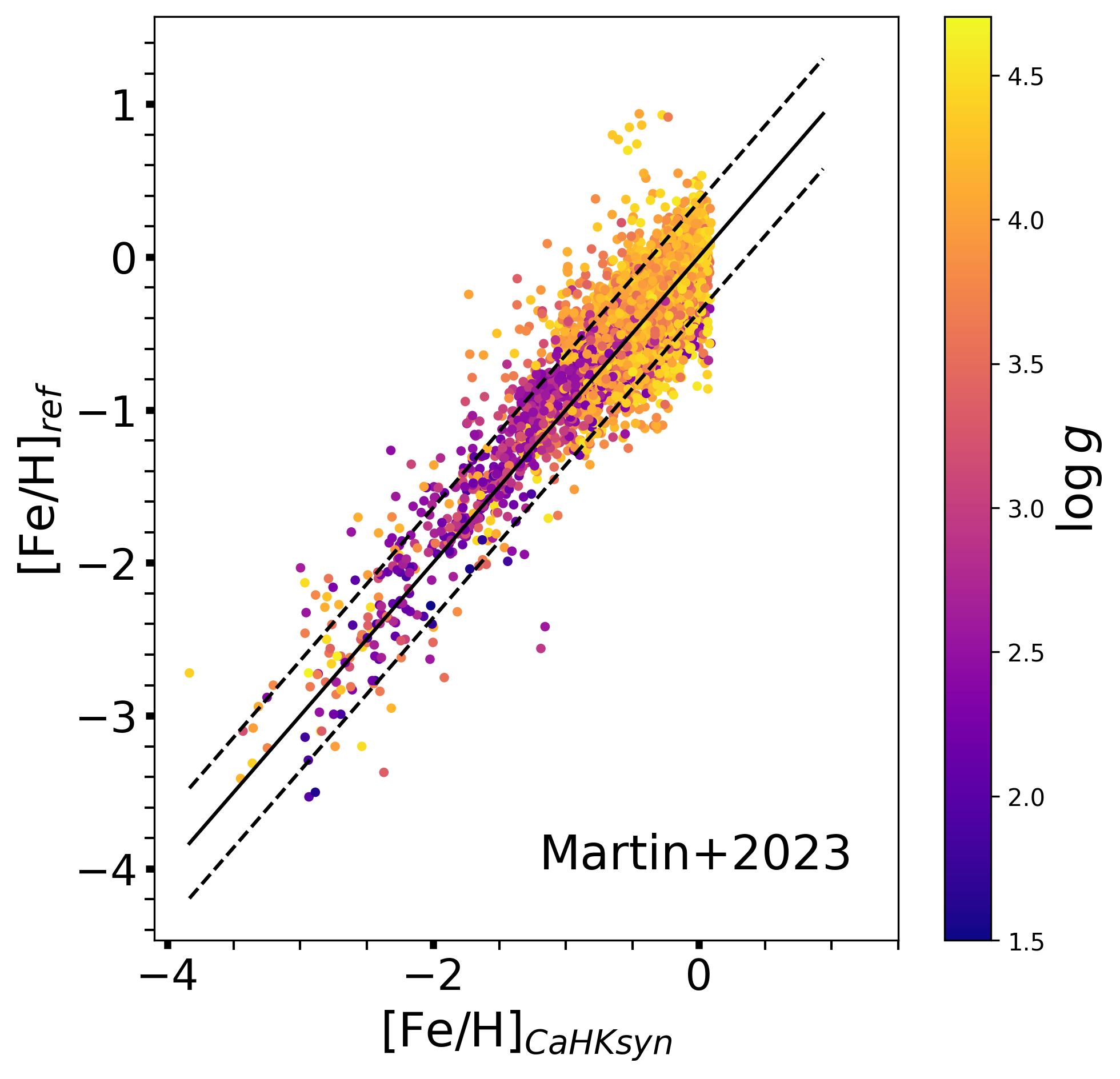}
    \end{minipage}
    \centering
    \begin{minipage}{0.29\linewidth} 
    \includegraphics[width=0.9\textwidth]{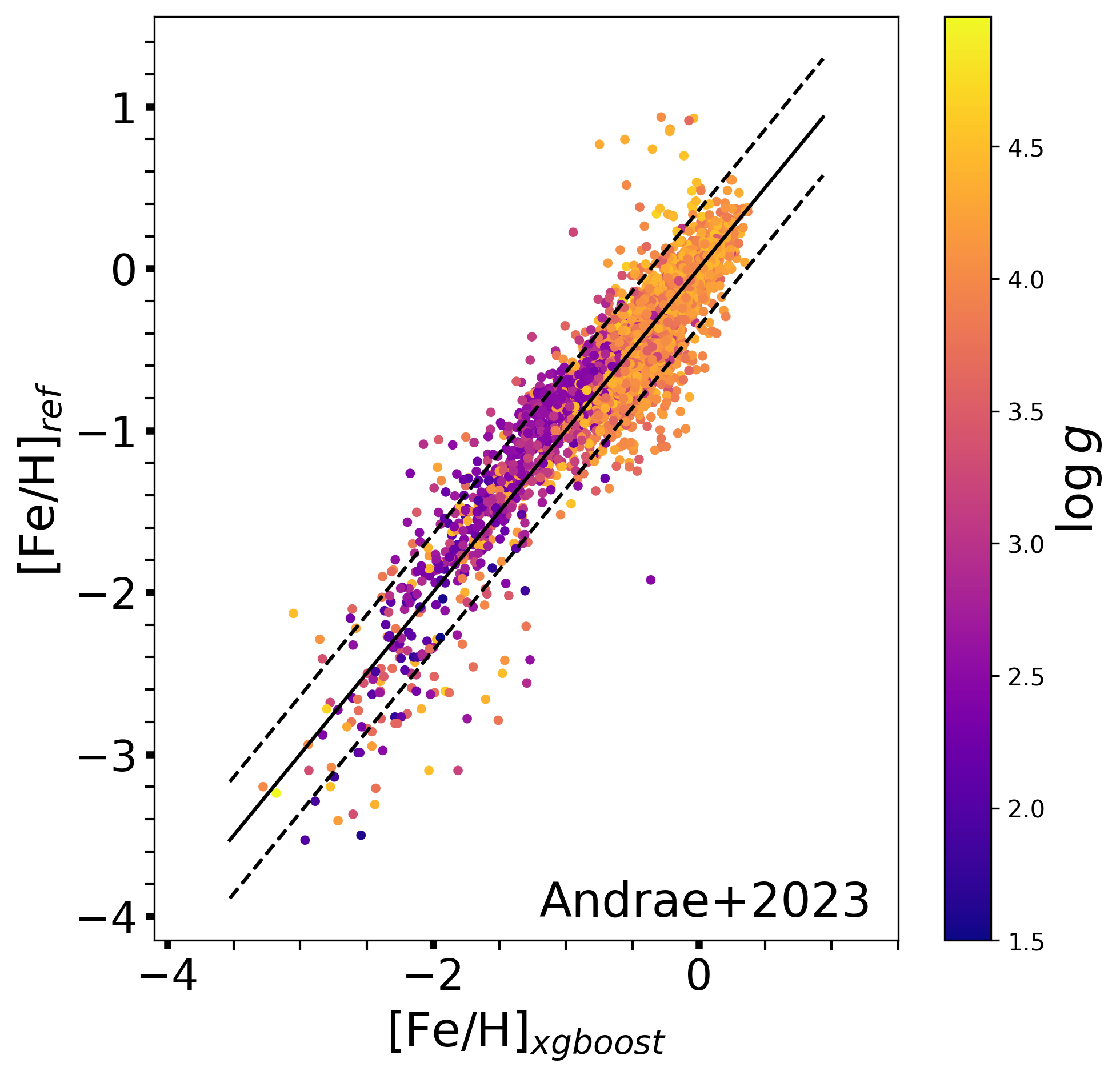}
    \end{minipage}
    \caption{Comparison of our derived metallicities with those from the \cite{2023ApJS..267....8A} (XGBOOST), and \cite{2023arXiv230801344M} ($\mathrm{CaHK}_{synth}$) catalogues. Top: from left to right the metallicities of our, XGBOOST, and $\mathrm{CaHK}_{synth}$ catalogues are plotted, respectively, for the GALAH-SAGA validation dataset ($\mathrm{[Fe/H]}_{ref}$). The color-coding reflects the effective temperature of the stars. Middle and bottom: as in the top panels, but here the color-coding depicts the color excess and surface gravity, respectively. The solid black line shows the 1-1 line, while the dashed lines show a $\sigma=$0.36 dex uncertainty.}
   \label{fig:cat_comp}
\end{figure*}

As already described in the introduction, many studies have taken advantage of the wealth of information encapsulated in the Gaia BP/RP spectra, and have provided to the community catalogues of stellar atmospheric parameters. Specifically, the catalogues of \cite{2023ApJS..267....8A} and \cite{2023arXiv230801344M} have been shown to work very well in the metal-poor regime. We used the GALAH-SAGA verification sub-dataset (Figure \ref{fig:feh_estimation}) to compare the metallicities we estimated versus those of \cite{2023ApJS..267....8A} and \cite{2023arXiv230801344M}. The $\mathrm{[Fe/H]}_{inf}$ we estimated for this sub-dataset are independent of the fitting procedure. Figure \ref{fig:cat_comp} shows the performance of each catalogue. At first glance it is visible that the catalogue of \cite{2023arXiv230801344M} performs better in the metal-poor regime than ours, and that of \cite{2023ApJS..267....8A}. However, the difference in accuracy of the inferred metallicities in all three catalogues is comparable, specifically for $\mathrm{[Fe/H]}_{ref}<-2$ the iron abundances of \cite{2023arXiv230801344M} and \cite{2023ApJS..267....8A} have $\sigma\sim0.39$ and are 0.1 dex better than ours. For $\mathrm{[Fe/H]}_{ref}<-3$ the standard deviation of the estimated metallicities in all three catalogues is the same, that is $\sim0.36$ dex. In the metal-rich regime, our metallicities have uncertainties that are $\sim0.2$ dex higher than those of the other two catalogues, whose performance is similar, $\sigma\sim0.24$ dex.

%
   

\section{Summary}
We applied the metal-poor star candidate selection recipe described in Paper I \citep{2022A&A...666A..58X} on $Gaia$ DR3 BP/RP spectra. In order to do so, we updated the selection method. Specifically, instead of using effective temperature and surface gravity information, we only used the flux-ratios - $fr\mathrm{_{G/CaNIR}}$ and $fr\mathrm{_{CaHK/H\beta}}$ - determined in Paper I to estimate the metallicity of stars. We addressed the extinction by means of dereddening the spectra before computing the flux-ratios, and found that the method can be applied to stars with color excesses $E(B-V)\leq1.5$. We then used BP/RP spectra through a cross-match between $Gaia$ DR3 and GALAH DR3, as well as with the SAGA database, to validate the selection method. We were able to estimate the $\mathrm{[Fe/H]}$, solely with the use of the flux-ratios, with an uncertainty of $\sigma_{\mathrm{[Fe/H]}_{inf}}\sim0.36$ dex. Next, we assessed to which degree OBA stars could contaminate a metal-poor candidate sample selected via the method described herein. We found that it is not very likely, as long as one has high level color excesses at their disposal to perform the dereddening of the spectra. Following, we selected stars from $Gaia$ DR3 via our updated selection procedure for spectroscopic validation. We observed 26 stars of which 100\% had $\mathrm{[Fe/H]}<-2.0$, 58\% had $\mathrm{[Fe/H]}<-2.5$ and 8\% had $\mathrm{[Fe/H]}<-3.0$. We inferred metallicites for this sample of stars prior to observations with an uncertainty $\sigma_{\mathrm{[Fe/H]}_{inf}}\sim0.31$. Finally, we assembled a catalogue of metallicities for 10 861 062, of which 225 498 have $\mathrm{[Fe/H]}_{inf}<-2.0$.

%
%

\begin{acknowledgements}
    We thank the anonymous referee for their comments, which helped improve this manuscript. This work was funded by the Deutsche Forschungsgemeinschaft (DFG, German Research Foundation) -- Project-ID 138713538 -- SFB 881 (``The Milky Way System'', subproject A04). This research was supported by the Australian Research Council Centre of Excellence for All Sky Astrophysics in 3 Dimensions (ASTRO 3D), through project number CE170100013. This work was supported by computational resources provided by the Australian Government through the National Computational Infrastructure (NCI) under the National Computational Merit Allocation Scheme and the ANU Merit Allocation Scheme (project y89). TXD acknowledges support from the Heidelberg Graduate School for Physics (HGSFP). TTH acknowledges support from the Swedish Research Council (VR 2021-05556). This work made use of the Third Data Release of the GALAH Survey (Buder et al. 2021). The GALAH Survey is based on data acquired through the Australian Astronomical Observatory, under programs: A/2013B/13 (The GALAH pilot survey); A/2014A/25, A/2015A/19, A2017A/18 (The GALAH survey phase 1); A2018A/18 (Open clusters with HERMES); A2019A/1 (Hierarchical star formation in Ori OB1); A2019A/15 (The GALAH survey phase 2); A/2015B/19, A/2016A/22, A/2016B/10, A/2017B/16, A/2018B/15 (The HERMES-TESS program); and A/2015A/3, A/2015B/1, A/2015B/19, A/2016A/22, A/2016B/12, A/2017A/14 (The HERMES K2-follow-up program). We acknowledge the traditional owners of the land on which the AAT stands, the Gamilaraay people, and pay our respects to elders past and present. This paper includes data that has been provided by AAO Data Central (datacentral.org.au).
This work has made use of data from the European Space Agency (ESA) mission
{\it Gaia} (\url{https://www.cosmos.esa.int/gaia}), processed by the {\it Gaia}
Data Processing and Analysis Consortium (DPAC,
\url{https://www.cosmos.esa.int/web/gaia/dpac/consortium}). Funding for the DPAC
has been provided by national institutions, in particular the institutions
participating in the {\it Gaia} Multilateral Agreement. This work has made use of the Python package $Gaia$XPy, developed and maintained by members of the $Gaia$ Data Processing and Analysis Consortium (DPAC) and in  particular, Coordination Unit 5 (CU5), and the Data Processing Centre located at the Institute of Astronomy, Cambridge, UK (DPCI).

\end{acknowledgements}

\bibliography{references}{}
    \citestyle{egu}
    \bibliographystyle{aa}
    
\begin{appendix}
\section{Additional figures.}
We present additional figures which are described in Section \ref{selection}. We show the inferred metallicity distribution of candidate metal-poor stars that are located on the flux-ratio plane between Cutoff2 and the more stringent cut we used to select stars for observations. Lastly, we show the metallicity distribution of metal-poor candidates below the stringent cut off that were not included in the target list.

\begin{figure}
    \centering \includegraphics[width=0.35\textwidth]{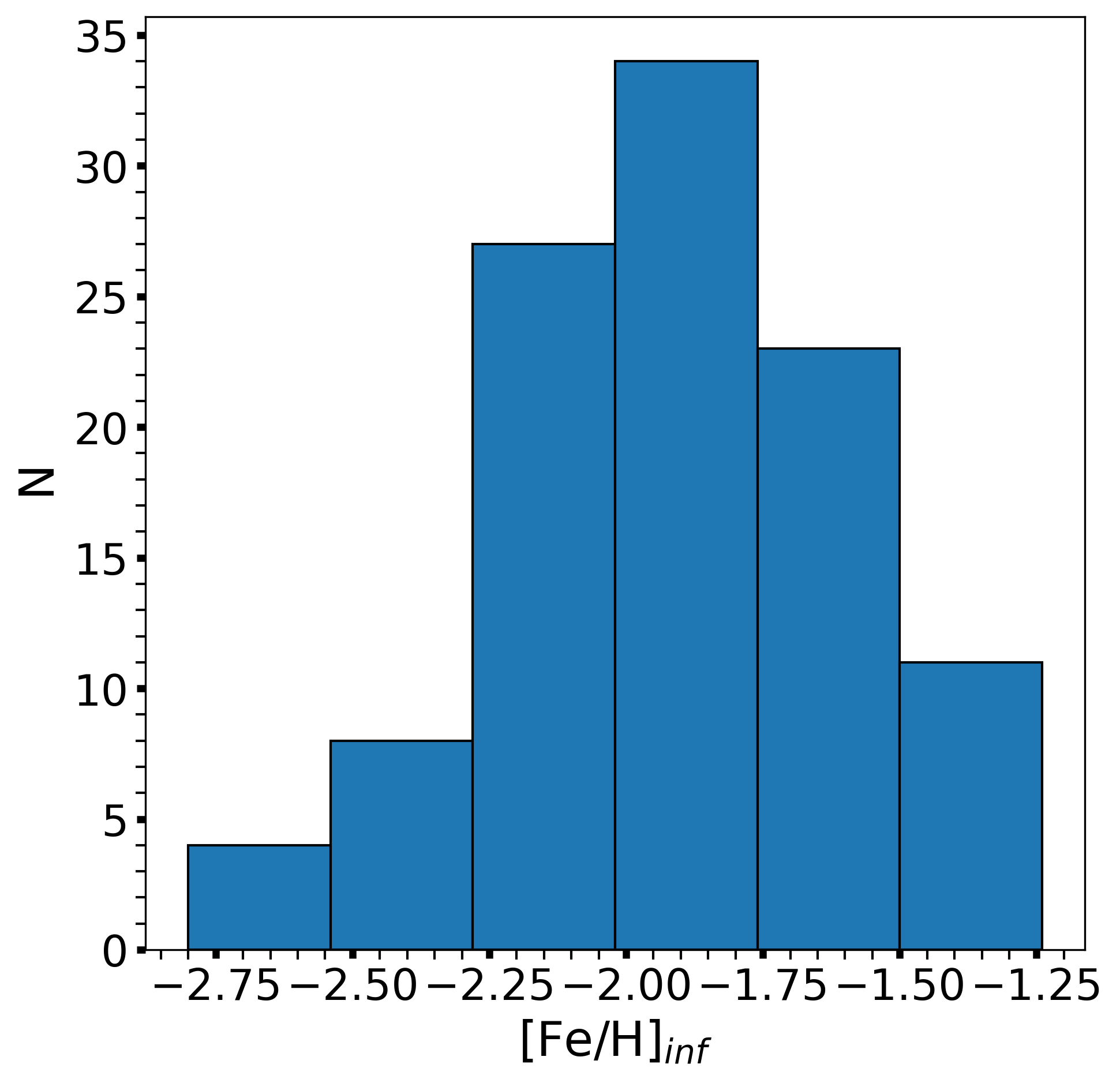}
    \caption{Distribution of $\mathrm{[Fe/H]}_{inf}$ for metal-poor candidates located between Cutoff2 and the cut off we used to select candidates for observations.}
    \label{fig:above_cut}
\end{figure}

\begin{figure}
    \centering \includegraphics[width=0.35\textwidth]{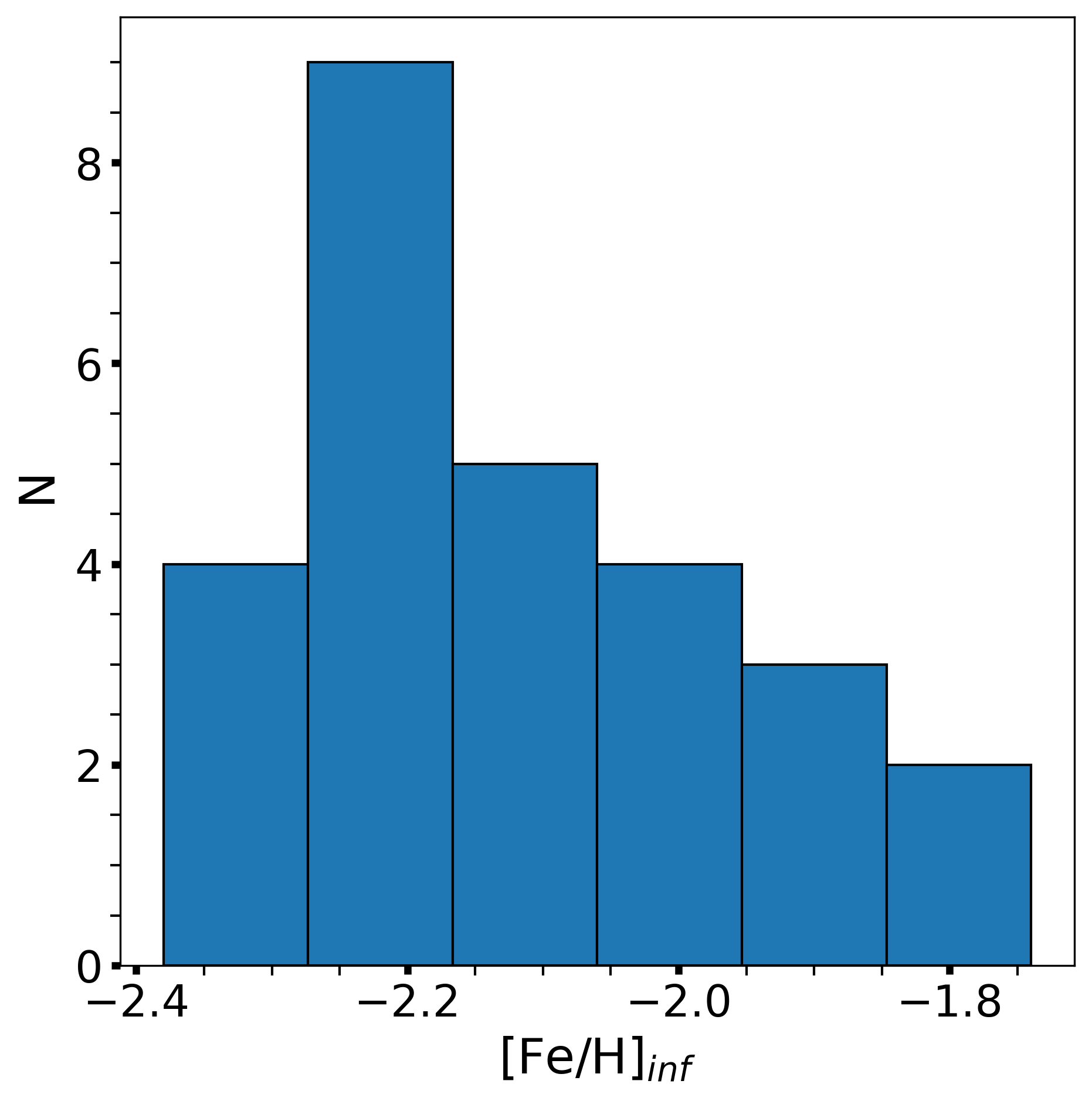}
    \caption{Distribution of $\mathrm{[Fe/H]}_{inf}$ of the 27 metal-poor candidates that were not included in the final target list.}
    \label{fig:not_included}
\end{figure}
\end{appendix}
\end{document}